\documentclass[preprint, 11pt]{aastex}
\usepackage{lscape}
\usepackage{graphicx}
\usepackage{epstopdf}

\newcommand{\masyr}{mas\,yr$^{-1}$}
\newcommand{\kms}{km\,s$^{-1}$}
\newcommand{\degree}{$^{\circ}$}
\newcommand{\pmra}{$\mu_{\alpha}$}
\newcommand{\pmdec}{$\mu_{\delta}$}
\newcommand{\ergcms}{erg cm$^{-2}$ s$^{-1}$}
\newcommand{\ergs}{erg s$^{-1}$}

\begin{document}

\slugcomment{Accepted to {\it ApJ} 2010 Feb 9}
\title{The Enigmatic Young Low-Mass Variable TWA 30\altaffilmark{1}}
\shortauthors{Looper et al.}
\shorttitle{TWA 30}

\author{Dagny L. Looper\altaffilmark{2,3}, 
Subhanjoy Mohanty\altaffilmark{4},
John J. Bochanski\altaffilmark{5}, 
Adam J. Burgasser\altaffilmark{3,5,6}, 
Eric E. Mamajek\altaffilmark{7}, 
Gregory J. Herczeg\altaffilmark{8},
Andrew A. West\altaffilmark{9},
Jacqueline K. Faherty\altaffilmark{3,10,11},
John Rayner\altaffilmark{2,3}, 
Mark A. Pitts\altaffilmark{2,3}, 
J. Davy Kirkpatrick\altaffilmark{12}}

\altaffiltext{1}{This paper includes data gathered with the 6.5-m Magellan Telescopes located at Las Campanas Observatory, Chile.}

\altaffiltext{2}{Institute for Astronomy, University of
Hawai'i, 2680 Woodlawn Dr, Honolulu, HI 96822, USA; dagny@ifa.hawaii.edu} 

\altaffiltext{3}{Visiting Astronomer at the Infrared Telescope Facility, which
is operated by the University of Hawaii under Cooperative Agreement
no. NCC 5-538 with the National Aeronautics and Space Administration,
Office of Space Science, Planetary Astronomy Program} 

\altaffiltext{4}{Imperial College London, 1010 Blackett Lab., Prince Consort Road, London SW7 2AZ, UK}

\altaffiltext{5}{MIT Kavli Institute for Astrophysics \&
Space Research, 77 Massachusetts Ave, Building 37-664B, Cambridge, MA
02139, USA}

\altaffiltext{6}{Center for Astrophysics and Space Science, University of California San Diego, 9500 Gilman Drive, Mail Code 0424, La Jolla, CA 92093, USA}

\altaffiltext{7}{Department of Physics and Astronomy, University of Rochester, P.O. Box 270171;
500 Wilson Boulevard, Rochester, NY 14627-0171, USA}

\altaffiltext{8}{Max-Planck-Institut fŸr extraterrestriche Physik, Giessenbachstra§e, 85748 Garching, Germany}

\altaffiltext{9}{Department of Astronomy, Boston University, 725 Commonwealth Ave, Boston, MA 02215, USA}

\altaffiltext{10}{Department of Physics and Astronomy, Stony Brook University, Stony Brook, NY 11794-3800, USA}

\altaffiltext{11}{Department of Astrophysics, American Museum of Natural History, Central Park West at 79th Street, New York, NY 10034, USA}

\altaffiltext{12}{Infrared Processing and Analysis Center, MS 100-22, California Institute of Technology, Pasadena, CA 91125, USA}

\begin{abstract}

TWA 30 is a remarkable young (7~$\pm$~3 Myr), low-mass (0.12~$\pm$~0.04 M$_{\sun}$), late-type star (M5~$\pm$~1) residing 42~$\pm$~2 pc away from the sun in the TW Hydrae Association.  It shows strong outflow spectral signatures such as [S II], [O I], [O II], [O III], and Mg I], while exhibiting weak H$\alpha$ emission ($-$6.8~$\pm$~1.2 \AA).  Emission lines of [S II] and [O I] are common to T Tauri stars still residing in their natal molecular clouds, while [O III] and Mg I] emission lines are incredibly rare in this same population; in the case of TWA 30, these latter lines may arise from new outflow material colliding into older outflow fronts.  The weak H$\alpha$ emission and 
small radial velocity shifts of line emission relative to the stellar frame of rest (generally $\lesssim$10 \kms) suggest that the disk is viewed close to edge-on and that the stellar axis may be inclined to the disk, similar to the AA Tau system, based on its temporal changes in emission/absorption line strengths/profiles and variable reddening (A$_V$~=~1.5--9.0).  The strong Li absorption (0.61~$\pm$~0.13 \AA) and common kinematics with members of the TWA confirm its age and membership to the association.  Given the properties of this system such as its proximity, low mass, remarkable outflow signatures, variability, and edge-on configuration, this system is a unique case study at a critical time in disk evolution and planet-building processes.  

\end{abstract}


\section{Introduction}

Over the past 30 years, a universal but complex paradigm of star formation has been developed through many observations and theoretical breakthroughs \citep[e.g.,\ ][]{1987ARA&A..25...23S, 1998apsf.book.....H}.  
Briefly, the stellar birth process begins with the gravitational collapse of material within giant molecular cloud complexes (Class 0 objects; \citealt{1987ApJ...312..788A}) and progresses as the embedded protostar accretes material from an enshrouding infall envelope (Class I), with this transition lasting $\sim$200 kyr \citep{2008ApJ...684.1240E}.  As the gas clears, the revealed protostar continues to accrete from a circumstellar disk of material (Class II, Classical T Tauri stars, cTTS; \citealt{1989A&ARv...1..291A}).  By $\sim$10 Myr, accretion ceases \citep{2009arXiv0911.3320F,2003MNRAS.342..876W}, and the object is now a pre--main sequence star, possibly with a debris disk (Class III and later) capable of assembling terrestrial planets over timescales of 10--100 Myr \citep{2004E&PSL.223..241C}.  This process from cloud core to pre--main sequence star is marked by a wide array of dynamic phenomena, including outflows and accretion.

Recent evidence suggests that the accretion processes at work in Class II systems operate in the regime of very low-mass stars and brown dwarfs as well (e.g., \citealt{2003ApJ...582.1109W,2005ApJ...626..498M}).  Accretion rates scale roughly as $\dot M \sim M^2$ \citep{2003ApJ...592..266M,2004A&A...424..603N,2004AJ....128.1294C,2005ApJ...626..498M,2006ApJ...639L..83A}, meaning lower mass systems should display weaker signatures of accretion.  Forbidden emission lines (FELs) in optical spectra, arising from low-density optically thin gas, have been used to identify a handful of outflows originating from very low-mass stars and brown dwarfs (e.g., \citealt{2001A&A...380..264F,2003ApJ...592..266M, 2004ApJ...616.1033L,2005ApJ...626..498M}).   Only five young brown dwarfs have had their outflows spatially resolved -- three optical jets (via spectroastrometry; \citealt{2005Natur.435..652W,2007ApJ...659L..45W,2009ApJ...691L.106W}) and two molecular outflows (via direct imaging; \citealt{2008ApJ...689L.141P,2005ApJ...633L.129B}).  Identification of such systems is hampered not only by their intrinsic faintness but also by the distances (d~$>$~120 pc) to the nearest star-forming regions (e.g., Sco-Cen, Taurus, Orion), making it difficult to resolve structure on the scales of disks ($\sim$10--100 AU) and jets (a few hundred AU).  Hence, progress in understanding the size, morphology, and energetics of jets powered by very low-mass stars and brown dwarfs is limited.

Fortunately, the recent identification of new low-mass members of the nearby TW Hydrae Association (TWA) may provide more ideal systems for studies of disk and jet structures in this mass regime.  Indeed, one of the three brown dwarfs with a resolved optical jet, 
2MASSW J1207334$-$393254 (2M1207AB, also known as TWA 27AB; \citealt{2002ApJ...575..484G,2004A&A...425L..29C}), is a member of the TWA.  The namesake of the TWA, TW Hydrae, was the first cTTS found in isolation \citep{1976ApJS...30..491H} -- located only 54~$\pm$~6 pc from the Sun \citep{2007A&A...474..653V}, 23 degrees above the Galactic Plane and far from any molecular cloud \citep{1978ppeu.book..171H,1983A&A...121..217R,2009PASJ...61..585T}.  More than a decade after the discovery of TW Hydrae, \cite{1989ApJ...343L..61D} and \cite{1992AJ....103..549G} found four more T Tauri stars in the same vicinity by using the IRAS Point Source Catalog and targeting stars with infrared excess.  Based on their common X-ray activity, \cite{1997Sci...277...67K} postulated that these five objects formed a physical association of young stars, which they termed the TW Hydrae Association.   Soon after, targeted surveys using ROSAT All-Sky Survey (RASS) data \citep{1999A&A...346L..41S,1999ApJ...512L..63W,2001ApJ...549L.233Z}, kinematic surveys \citep{2003ApJ...599..342S,2005A&A...430L..49S}, and photometric near-infrared (NIR) surveys \citep{2002ApJ...575..484G,2007ApJ...669L..97L} discovered several more members, bringing the total to 23 confirmed systems (\citealt{2005ApJ...634.1385M}), of which five contain brown dwarfs.  

These systems share similar kinematics and indicators of youth such as strong chromospheric activity and, for late-K to M type members, Li I $\lambda$6708 absorption.  Isochronal ages derived from the Hertzsprung-Russell diagram yield age estimates of $\sim$8 Myr for the TWA \citep{1998ApJ...498..385S,1999ApJ...512L..63W,2000ApJ...530..867W,2006A&A...459..511B}.  At an average distance of 53 pc (Mamajek, in preparation), the TWA is the nearest association containing actively accreting young stars\footnote{Although, a recently identified young brown dwarf, 2MASS J0041353$-$562112, shows signs of accretion and may be a possible $\beta$ Pic or Tuc-Hor member, placing it at $\sim$35--50 pc \citep{2009ApJ...702L.119R}.}, making it an attractive target for studying planet formation and the evolution of circumstellar disks.  Moreover, members of the TWA span a range of disk evolutionary stages, with some members showing signs of actively accreting disks, passive/non-accreting disks, or debris disks,  while other members show no signs of circumstellar material \citep[e.g., ][]{1999ApJ...521L.129J,2005ApJ...631.1170L}.  

We have undertaken a survey to identify additional low-mass members to the TWA, which could provide important new case studies of accretion and outflow in substellar-mass objects residing at close distances ($\sim$50 pc).  In this paper, we report the discovery of a new, low-mass member of the TWA, 2MASS J11321831$-$3019518, which we term TWA 30.  Its membership is confirmed by its kinematics, Li I $\lambda$6708 absorption strength, and signatures of a low-gravity photosphere.  TWA 30 has strong forbidden lines of [O I], [O II] and [S II] within 30 \kms\ of the stellar rest velocity, possibly indicating an outflow from a nearly edge-on system.  The presence of [O III] and Mg I] in the spectrum of TWA 30 are extremely rare for a cTTS and have not been observed before in a young system residing outside of its natal molecular cloud.  The fluctuating continua levels from both the optical and the NIR spectra suggest that they may be affected by highly variable reddening on timescales of a day to several weeks.  In $\S$ 2, we describe the discovery and observations of TWA 30; in $\S$ 3, we analyze the kinematics, spectral morphology, emission/absorption lines, X-ray activity, and estimate an age and mass of TWA 30; in $\S$ 4, we discuss evidence for this system having an inclined stellar axis to an edge-on disk and update the disk fraction of the TWA; finally in $\S$5, we give our conclusions.

\section{Observations}

\subsection{Discovery}

\begin{figure*}[htbp]
\centering
\includegraphics[scale=0.4]{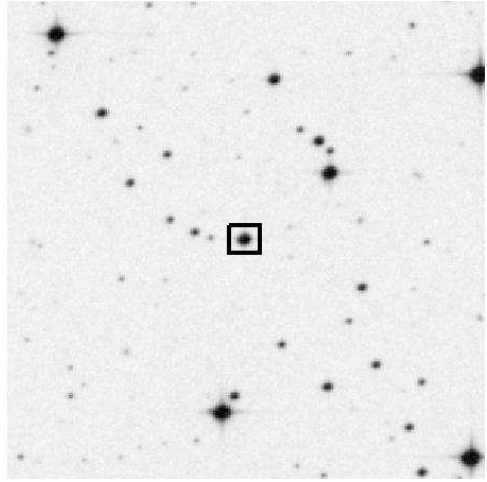}
\caption{Finder chart for TWA 30.  The field is centered on the target and is 5 arcmin on a side, with north up and east to the left.  The image is in the $R$-band from the DSS II (epoch 1991 Feb 22 UT).  A box of 20$\arcsec$ on a side marks the location of the target.}
\label{fig:finder}
\end{figure*}

We searched for new members of the TWA that appear in the RASS X-ray Faint Source Catalog \citep{2000IAUC.7432....3V} over the right ascension (RA) range 10$^h.5 -12^h.5$ and declination (DEC) range $-$25\degree~to $-$40\degree, with magnitudes of 7.5~$<J<$~10 in the Two Micron All-Sky Survey (2MASS) Point Source Catalog (PSC; \citealt{2003yCat.2246....0C, 2006AJ....131.1163S}).  Previous searches of RASS data for TWA members \citep{1999ApJ...512L..63W,2001ApJ...549L.233Z} have relied on the X-ray Bright Source Catalog \citep{1999A&A...349..389V}, with the exception of the study by \cite{1999A&A...346L..41S}, which examined faint sources in RASS data over a much smaller field of view than our survey.  We identified TWA 30 in our sample as a faint X-ray source (1RXS J113217.7$-$302007) with (2.47~$\pm$~0.96) $\times$ 10$^{-2}$ counts s$^{-1}$, $J$~=~9.64, and $J-K_s$~=~0.88.  TWA 30 is located in the southwest region of the TWA (a finder chart for it is shown in Figure \ref{fig:finder}).  Further details of our search methodology, spurious candidates, and other new members will be reported in a forthcoming paper.  

This source was independently identified by \cite{2003ApJ...599..342S} as SSS 113218$-$3019 on the basis of unpublished optical spectral data.  They deduced that it is distant ($>$100 pc), and hence likely a member of the Lower Centaurus-Crux region.  In the following sections we show evidence that it is nearby and a member of the TWA.

\subsection{Optical Spectroscopic Data}

\subsubsection{MagE Spectroscopy}
\begin{figure*}[htbp]
\centering
\includegraphics[scale=0.8]{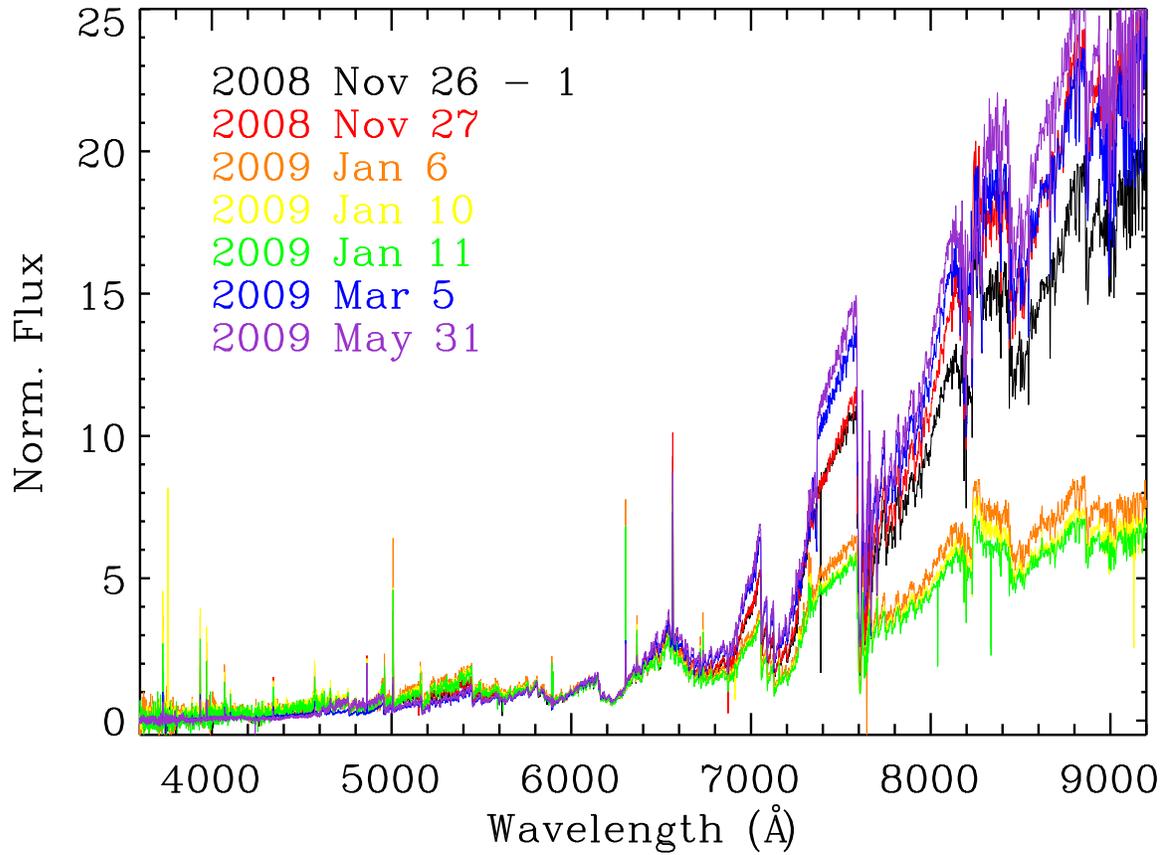}
\caption{Optical spectra of TWA 30 taken with MagE at the seven indicated epochs.  Three spectra were taken sequentially on 2008 Nov 26 UT, and we only show the first one here as all three are identical.  The spectra have been normalized at 6000 \AA, showing the highly variable nature of this object in both emission lines and the amount of reddening present.}
\label{fig:all_opt_long}
\end{figure*}

TWA 30 was observed nine times with the Magellan Echellete \citep[MagE;][]{2008SPIE.7014E.169M} on the 6.5 m Clay Telescope at Las Campanas Observatory.  The details of each observation are recorded in Table \ref{table:log}.  MagE is a medium resolution, cross--dispersed echellette, covering the optical regime from 3,000 to  10,500 \AA .   The 0$\farcs$7 slit was employed for each observation, corresponding to  $R~\equiv~\lambda$ / $\Delta\lambda~\approx$~4100 with no binning.  A flux standard was observed during each run, and ThAr arc calibration images were obtained after each science exposure.  The spectrum was reduced with the MASE reduction tool pipeline \citep{2009PASP..121.1409B}.  Briefly, MASE is written in the Interactive Data Language (IDL) and incorporates the entire reduction and calibration process, including bias subtraction, flat fielding, wavelength and flux calibration.  Typical wavelength solutions are accurate to $\sim$5--7 km s$^{-1}$.  The reduced spectrum from 2009 Jan 10 UT is shown in Figures \ref{fig:mage_spec1}, \ref{fig:mage_spec2}, and \ref{fig:mage_spec3}.  Three of the spectra were taken on the same night (2008 Nov 26 UT) in sequential exposures and are identical to one another; we show the first spectrum of this series and the other six spectra from different epochs in  Figure \ref{fig:all_opt_long}.

\subsubsection{MIKE Spectroscopy}
High resolution spectroscopy was obtained with the Magellan Inamori Kyocera Echelle \citep[MIKE;][]{2003SPIE.4841.1694B} on 2008 Nov 27 UT.   TWA 30 was observed with the red camera and the 0$\farcs$7 slit,  with coverage from 4,900 to 9,200 \AA\ and $R \sim$ 24,000.  Binning was set at 2 $\times$ 2.  ThAr arcs were obtained after the science exposure, and a flux standard was observed during the night for relative flux calibration.  The spectrum was reduced using the MIKE IDL pipeline\footnote{The pipeline is available at http://web.mit.edu/\~{}burles/www/MIKE/.}, producing a calibrated 1D spectrum with wavelength solutions accurate to $\sim$1--2 \kms.

\subsection{NIR Spectroscopic Data}

We observed TWA 30 eleven times with the SpeX spectrograph \citep{2003PASP..115..362R} on the 3.0 m NASA Infrared Telescope Facility (IRTF) in three modes with the 0$\farcs$5 slit: short cross-dispersed (SXD: 0.8--2.4 $\mu$m; R $\sim$ 1200), long cross-dispersed (LXD: 1.9--4.2 $\mu$m; R $\sim$1500), and prism (0.7--2.5; R $\sim$ 150).  Details of these observations are listed in Table \ref{table:log}.  Observations of a nearby A0 V star for telluric correction were made immediately after each observation along with internal flat-field and argon arc lamp frames for calibrations.  Reductions were carried out using the Spextool package version 3.4 \citep{2004PASP..116..362C,2003PASP..115..389V}.  The LXD data were first combined at each nod position; reductions using the Spextool package were otherwise standard.  

\subsection{Imaging}

We observed TWA 30 at the Cerro Tololo 4.0 m Blanco telescope on 2008 Dec 13 UT as part of an ongoing brown dwarf astrometric program.  The Infrared Side Port Imager (ISPI; \citealt{2004SPIE.5492.1582V}) was used along with the $J$-band filter.  The detector has a $\sim$10 arcmin field of view with a 0$\farcs$3 pixel plate scale.  Three images of 10 s exposures using 3 co-adds were obtained at each of the three dither positions.  Dark frames and domeflats were obtained at the start of each evening.   The raw images were median-combined to produce sky frames, which were then subtracted from the raw data.  The subsequent reduction procedures were based on the prescriptions put together by the ISPI team\footnote{See http://www.ctio.noao.edu/instruments/ir\_instruments/ispi/.}, utilizing a combination of IRAF routines as well as publicly available software packages: WCSTOOLS and SWARP.  Each image was flat-fielded, corrected for bad pixels, and flipped to orient N up and E to the left using the IRAF routine $\it{osiris}$ in the $\it{cirred}$ package.  Individual point sources were selected in each image using the IDL routine $\it{find}$ and then input into the WCSTOOLS task $\it{imwcs}$, which matches stars to the 2MASS PSC.  Once an initial world coordinate system (WCS) was set, we used the IRAF routine $\it{ccmap}$ to correct for the distortion prominent across the ISPI detector.   The final WCS residuals against 2MASS were $\sim$0.1 pixels in both the X and Y axes.  We used the software package $\it{swarp}$ to shift and add the 3 reduced dither positions and then used this final science frame to perform the astrometry.   Final RA, DEC coordinate positions of 11h 32m 18.19s, $-$30\degree\ 19$^{\prime}$ 51.8$\arcsec$ (equinox J2000) were determined using the IRAF routine $\it{wcstran}$, which combines the standard WCS parameters with the higher order distortion corrections calculated with $\it{ccmap}$.

\section{Analysis}

\subsection{Kinematics}

TWA 30 has proper motion estimates in the USNO-B1.0 catalog \citep{2003AJ....125..984M} and the SSS catalog \citep{2001MNRAS.326.1315H,2001MNRAS.326.1295H,2001MNRAS.326.1279H} of $\mu_{\alpha}$, $\mu_{\delta}$ of $-$86~$\pm$~5, $-$24~$\pm$~11 and $-$81~$\pm$~9, $-$28~$\pm$~9 \masyr, respectively.  It is also listed in the UCAC3 catalog \citep{2010arXiv1002.0556F} as 3UCAC 120-147008 with a proper motion of $\mu_{\alpha}$, $\mu_{\delta}$ of $-$89.6~$\pm$~1.3, $-$25.8~$\pm$~1.3 \masyr.  We adopt this latter measurement, which has the smallest errors, as the proper motion for TWA 30.

We determined the radial velocity (RV) of TWA 30 by measuring the line centers of H$\alpha$ and photospheric absorption lines in the MIKE spectrum (Li I $\lambda$6708, K I $\lambda$7699, and Na I $\lambda$8183)\footnote{We follow the convention of referring to lines in the optical regime at their rest velocity in air and round to the nearest integer.}, yielding a heliocentric RV of 12.3~$\pm$~1.5 \kms\ (Table \ref{table:properties}).  The uncertainty was calculated as the 1$\sigma$ standard deviation in the measurements.   To examine the RV variation over time, we calculated RVs from the nine lower resolution MagE spectra by cross-correlating each spectrum with the M6 SDSS template from \cite{2007AJ....133..531B} using the \textit{xcorl} package in IDL \citep{2003ApJ...583..451M,2009ApJ...693.1283W}.  Our analysis included the wavelength region from 6600 \AA\ to 7200 \AA\ that encompasses the TiO molecular bandhead that begins at 7056 \AA.  We also calculated the relative radial velocities by cross-correlating each MagE spectrum with the first spectrum taken chronologically on 2008 Nov 26 UT.  These RV values indicate a relative uncertainty of $\sim$2--3 \kms .  They are consistent with each other except for the spectrum taken on 2009 March 4 UT (and possibly 2009 May 31 UT), which shows a statistically smaller radial velocity than observed at other epochs (see Table \ref{table:rvs}).

Since no parallax measurement of TWA 30 is available, we used the convergent point analysis of \cite{2005ApJ...634.1385M} to estimate its distance.  For these calculations, we adopt the mean velocity of the TWA from Mamajek, in preparation based on revised {\it Hipparcos} astrometry \citep{2007A&A...474..653V} for TWA 1, 4, and 11, as well as the recent ground-based astrometric analyses of 2M1207AB (TWA 27AB; \citealt{2007ApJ...669L..45G,2007ApJ...669L..41B,2008A&A...477L...1D}).  The adopted TWA velocity is (\textit{U, V, W}) = ($-$10.2, $-$18.3, $-$4.9) \kms, with uncertainties of $\pm$~0.5~\kms\, in each component. The intrinsic 1D velocity dispersion in the TWA is approximately 0.8\,\kms\ \citep{2005ApJ...634.1385M}.

Most of the proper motion of TWA 30 is moving towards the TWA convergent point with near zero peculiar motion (1.2 \kms), 
a strong indication of membership. The proper motion of TWA 30 towards the convergent point is consistent with a cluster parallax of 23.8\,$\pm$\,1.2 mas or a distance of 42\,$\pm$\,2 pc.  This is much closer than the $>$100 pc spectrophotometric distance quoted by \cite{2003ApJ...599..342S}.  Given the highly variable nature of this object (see $\S$3.2 \& $\S$3.3), the kinematic distance is more reliable.  The mean distance to TWA members is 53 pc (Mamajek, in preparation), with individual members having distances between $\sim20-80$ pc.  

We conclude that both the radial velocity and proper motion are statistically consistent with TWA 30 being a TWA member.  Using the stated distance estimate, we derive the inferred tangential velocity and the (\textit{U,V,W}) space motion for TWA 30 with respect to the Sun (see Table  \ref{table:properties}).  For the latter we used the procedure of \cite{1987AJ.....93..864J} with updated galactic coordinate transformations from \cite{1989A&A...218..325M} and followed the convention that \textit{U} is positive towards the Galactic Center ({\it l}=0,{\it b}=0).  

\subsection{Optical Spectroscopy}

\subsubsection{Spectral Morphology}

We spectroscopically classified and determined the $A_v$ of each epoch of optical data from comparison to M3--M5.75 $\eta$ Cha templates \citep{2004ApJ...609..917L}.  This association has a similar age ($\sim$7 Myr; \citealt{1999ApJ...516L..77M,2001MNRAS.321...57L}) to the TWA and should therefore provide a better continuum fit with comparable surface gravities than would comparison to field templates, which have higher surface gravities.  These templates are: $\eta$ Cha 6 (M3), $\eta$ Cha 12 (M3.25), $\eta$ Cha 5 (M4), $\eta$ Cha 9 (M4.5), $\eta$ Cha 14 (M4.75), $\eta$ Cha 17 (M5.25), $\eta$ Cha 18 (M5.5), and $\eta$ Cha 16 (M5.75).   Both template and data were normalized at 7500 \AA, which was chosen as a normalization point due to the small amount of absorption at this wavelength \citep{1991ApJS...77..417K}.  Then the data were dereddened by increments of $A_v$=0.1 using the reddening law of \cite{1989ApJ...345..245C} (updated in the near-UV by \citealt{1994ApJ...422..158O})  to provide the best match to the template spectrum over the wavelength range 6000--8000 \AA.  We set the ratio of total-to-selective extinction R$_V$~$\equiv$~A$_V$ / E(B$-$V) to 3.1 (the standard for the ISM), although large grain sizes would have higher R$_V$ values (e.g., \citealt{2001ApJ...548..296W}).  The spectral features over this regime cover many broadband features typically used to classify M dwarfs such as TiO, VO, and CrH \citep{1991ApJS...77..417K}.   The best fit to the continuum, by eye, was then chosen, which simultaneously determined the spectral type and $A_v$.

\begin{figure*}[htbp]
\centering
\includegraphics[scale=0.7]{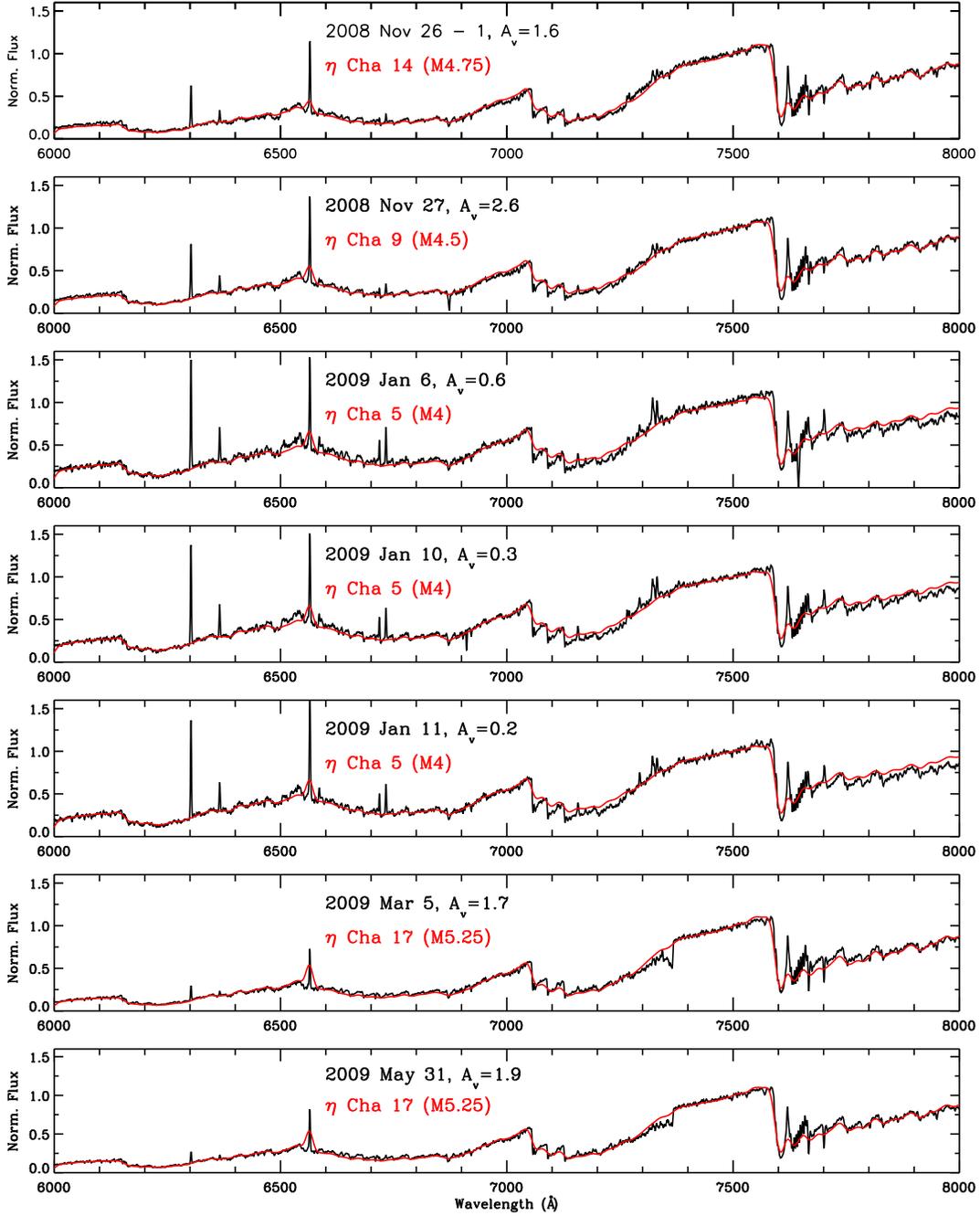}
\caption{Overlays of the MagE optical spectra of TWA 30 (black; R$\approx$4100) at each stated epoch with optical $\eta$ Cha templates (red; R$\approx$1000, \citealt{2004ApJ...609..917L}), which provided the closest match for each epoch (see $\S$3.2.1).  The spectra have been dereddened by the stated amount to provide the best fit.  All spectra have been normalized at 7500 \AA.}
\label{fig:opt_spt_matches}
\end{figure*}

The optimal fit of each epoch is overlaid in Figure \ref{fig:opt_spt_matches}.  The spectral type fits of the seven epochs varied between M4 and M5.25, which span timescales of hours to weeks between epochs.  The three spectra obtained on 2008 Nov 26 UT have a best fit to the M4.75 template, the spectrum obtained on 2008 Nov 27 UT has a best fit to an M4.5 template; the 2009 Jan 6, 10, and 11 UT spectra have a best fit to an M4 template; and the 2009 Mar 5 and 2009 May 31 UT have a best fit to an M5.25 template.   We discuss the possible underlying cause for these changes in spectral type, M4--M5.25, in $\S$4.2.

\subsubsection{Emission and Absorption Lines}

\begin{figure*}[htbp]
\centering
\includegraphics[scale=0.95]{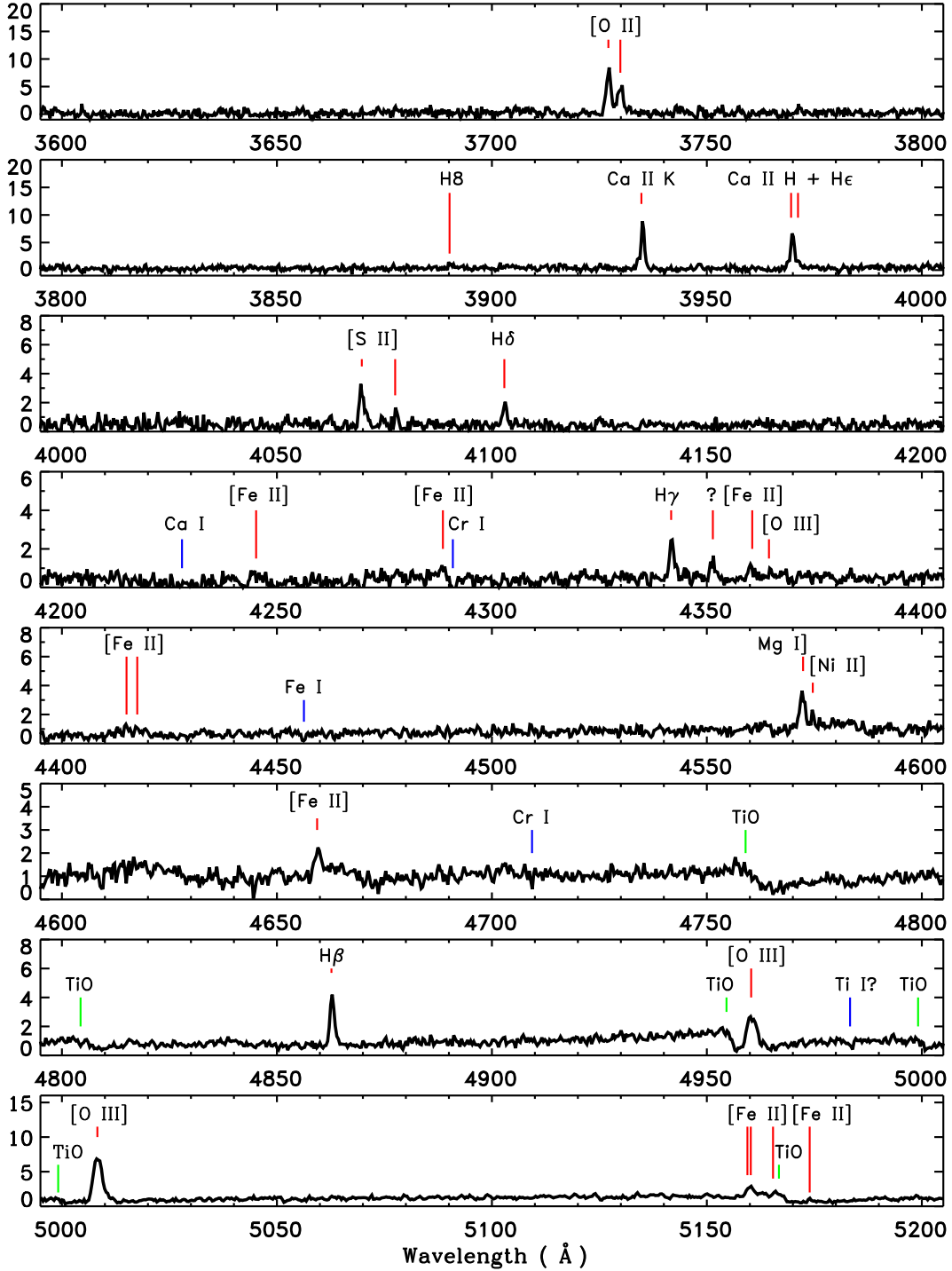}
\caption{The optical spectrum of TWA 30, obtained with the MagE spectrograph on 2009 Jan 10 UT.  Chromospheric emission lines and forbidden lines are indicated in red, absorption lines in blue, and molecular bands in green.  Many forbidden lines typically associated with outflows (i.e., [O I], [O II], [S II]) are seen, along with strong Balmer line and Ca II emission, typical for stars with an active accretion disk.  The spectral type derived from this observation is M4.}
\label{fig:mage_spec1}
\end{figure*}

\begin{figure*}[htbp]
\centering
\includegraphics[scale=0.95]{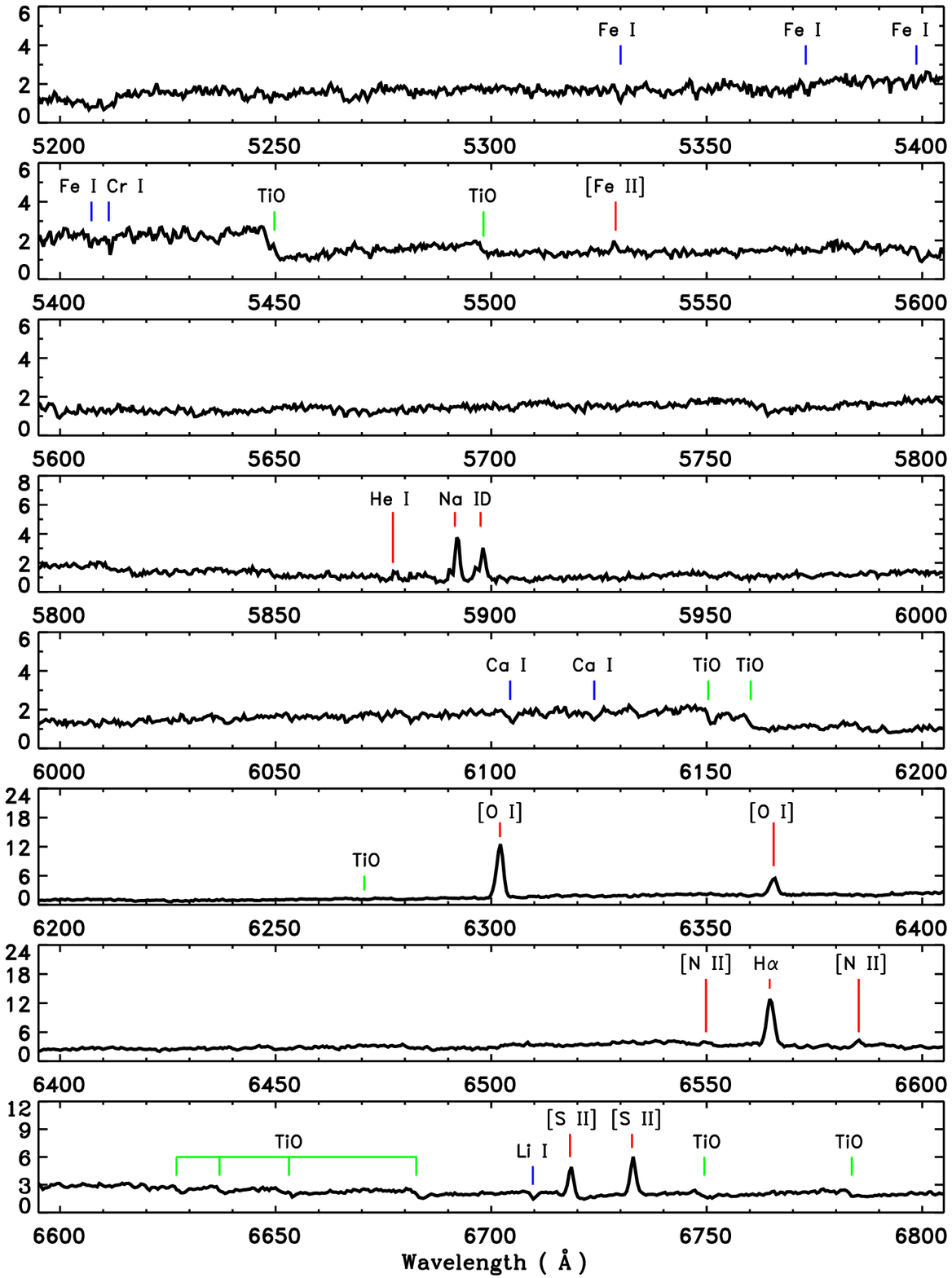}
\caption{Same as Figure \ref{fig:mage_spec1}.}
\label{fig:mage_spec2}
\end{figure*}

\begin{figure*}[htbp]
\centering
\includegraphics[scale=0.95]{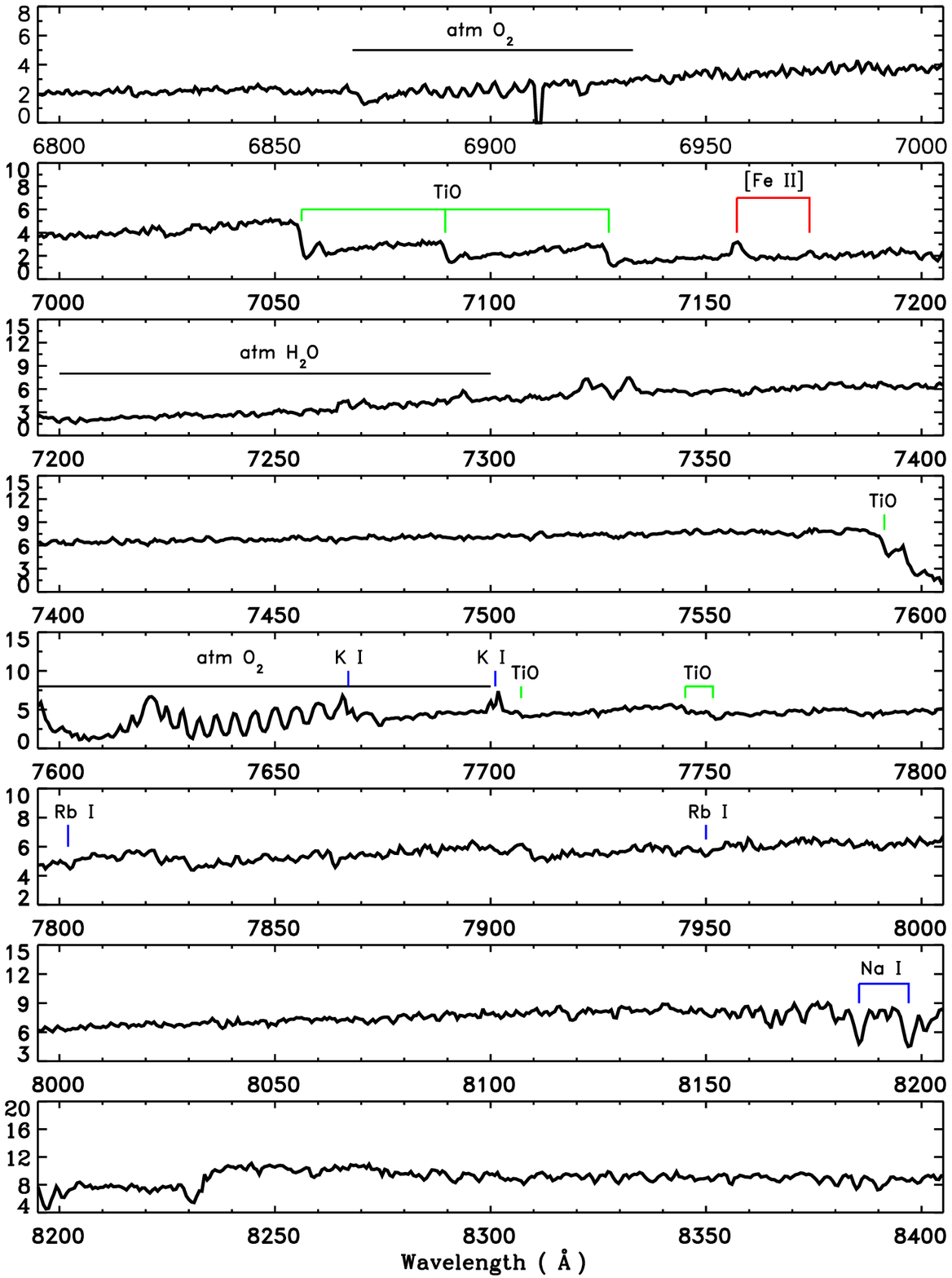}
\caption{Same as Figure \ref{fig:mage_spec1}.}
\label{fig:mage_spec3}
\end{figure*}

The alkali line strengths of Na I and K I, which are sensitive indicators of surface gravity \citep{1999ApJ...525..466L,1999AJ....118.1005M,2006ApJ...639.1120K,2009AJ....137.3345C}, are weaker in all of the optical spectra of TWA 30 than in the template dwarf spectra, indicating that this object has a lower surface gravity and is therefore younger than field dwarfs of similar spectral type.  However, K I $\lambda$7699 goes into emission in the 2009 Jan epochs, introducing some uncertainty in using it as a low gravity indicator.  We measured the equivalent widths (EW) of these lines along with many permitted and forbidden emission features\footnote{These lines were identified using the Atomic Line List v2.04 maintained at http://www.pa.uky.edu/$\sim$peter/atomic/ and from \cite{2009ApJS..182...33K} and references there-in.} over our seven epochs of MagE data and report these in Tables \ref{table:EWs_balmer}, \ref{table:EWs_notbalmer}, and \ref{table:EWs_abs} and show them in Figures \ref{fig:mage_spec1}, \ref{fig:mage_spec2}, and \ref{fig:mage_spec3}.  

We report the average over all nights of the Li I absorption and the H$\alpha$ emission EW, as these are indicators of age and accretion/chromospheric activity, respectively.   The prominent Balmer emission lines are a sign of youth in low--mass stars \citep[][and references therein]{2008AJ....135..785W}.  Furthermore, Li is fully depleted within $\sim$100 Myr for stars later than spectral type mid-K \citep{2004ARA&A..42..685Z},  placing an upper limit on the age of TWA 30.   It should be noted that for substellar objects with M~$<$~0.059--0.062 M$_{\sun}$, their core temperatures never become hot enough to destroy Li \citep{1993ApJ...413..364N}.  The Li EW of TWA 30 (0.61~$\pm$~0.13 \AA) is comparable to other TWA M dwarf members (0.40 to 0.65 \AA) and stronger than $\beta$ Pic M dwarf members (Li EW $<$ 0.4 \AA, age $\sim$12 Myr; \citealt{2004ARA&A..42..685Z}).

The FELs of [O I] and [S II] identified in Figures \ref{fig:mage_spec1} and \ref{fig:mage_spec2} are present in the spectra of many cTTS and are signs of an outflow and accretion \citep{2003ApJ...592..266M}.  These lines are typically associated with [N II] emission, which is only weakly present at 6548 and 6583 \AA.  We have also identified the high ionization species [O III] and [Fe III], which typically arise in planetary nebulae or winds from the most massive stars.  They are rarely seen in cTTS but are detected in the two high accretion rate cTTS: DG Tau (K5, \citealt{1995ApJ...452..736H}; \citealt{1996RMxAA..32..161R}) and XZ Tau B (M1, \citealt{2001ApJ...556..265W}).  The emission of the high-excitation [OIII] line requires extremely high temperatures and low densities, indicative of collisional shock fronts -- e.g., bow shocks formed by the jet running into the surrounding ambient medium \citep{1983ARA&A..21..209S,1989ApJ...339..987H,1995Ap&SS.233...11B,1996RMxAA..32..161R}.  Given the general lack of ISM gas in TWA \citep{2009PASJ...61..585T}, this medium may consist of older outflow material from TWA 30, which is overtaken and shocked by the fast moving jet.

Similarly, Mg I] $\lambda$4571 is clearly present in our spectra but is rarely seen in the spectra of cTTS.  \cite{2008ApJ...688..398E} have predicted the presence of Mg I] arising from gas in a circumstellar disk but noted that the detection may be difficult due to the strong continuum at these wavelengths.  Mg I] has previously been detected in XZ Tau B \citep{2001ApJ...556..265W} and the EXor pre-main sequence star VY Tauri (M0; \citealt{1990ApJ...360..639H}), which resides in the Taurus-Auriga Cloud.  

While TWA 30 has many strong spectral signatures indicating a strong outflow and hence an actively accreting circumstellar disk, it lacks any detectable He I emission at 4471, 6678, and 7065 \AA\ and has weak emission at 5876 \AA\ (see Table \ref{table:EWs_notbalmer}).  He I emission lines are seen in the accreting M4.75--M8 Taurus members in the sample of  \cite{2008ApJ...681..594H} and in the accreting late-M TWA members TWA 27AB (2M1207AB) and TWA 28 \citep{2009ApJ...696.1589H}.  He I is only weakly present in the accreting late-K TWA member TWA 1 and the early-M TWA 3A \citep{2009ApJ...696.1589H}.  He I emission in cTTS is thought to arise both in the infalling funnel flow (broad component) and in the accretion shock region on the stellar surface (narrow component).  We propose a scenario for the absence/weakness of the He I lines in our spectra in $\S$4.2.

\begin{figure*}[htbp]
\centering
\includegraphics[scale=0.85]{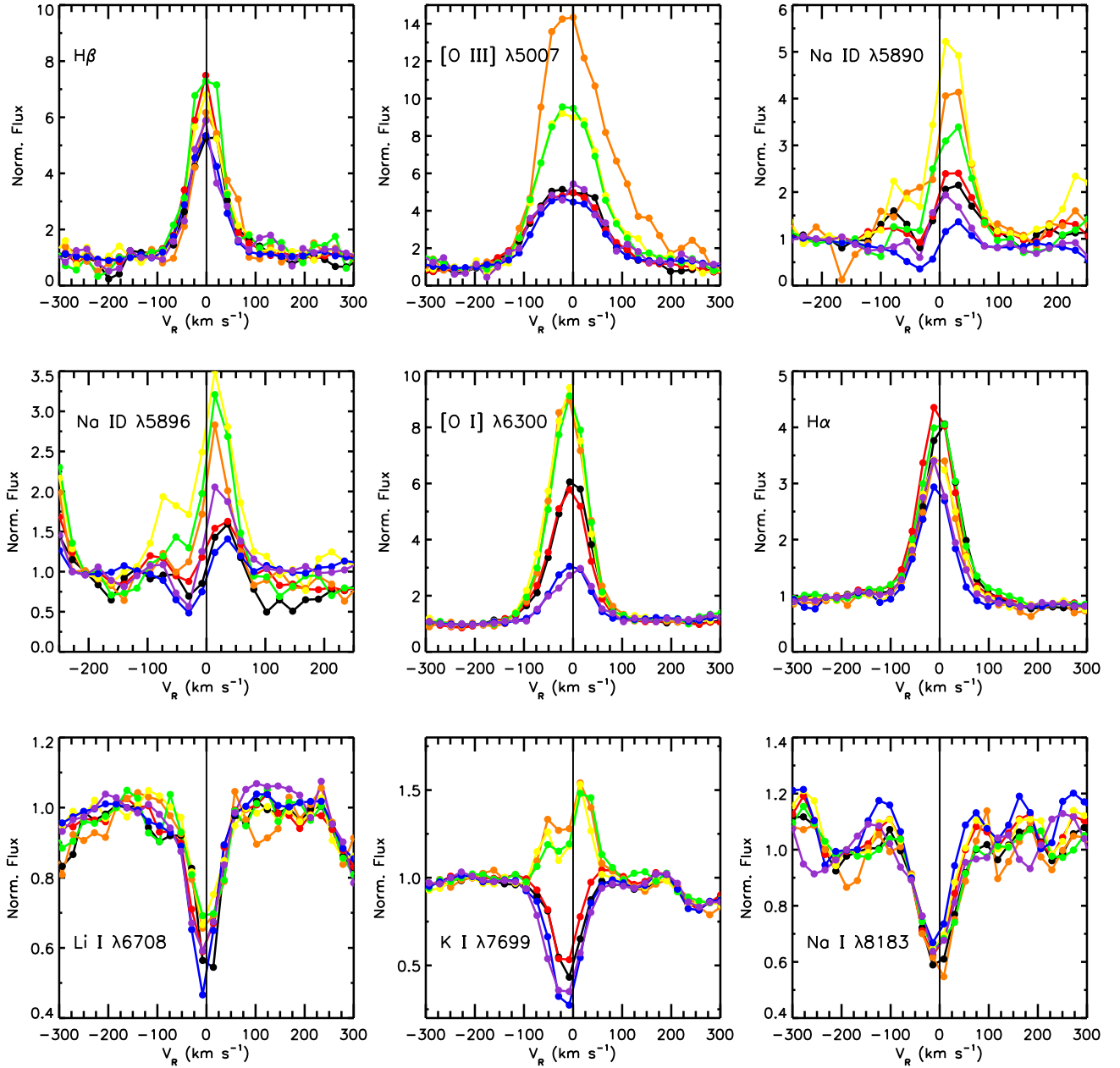}
\caption{The radial velocity profiles of several prominent features in the seven MagE spectra, color-coded identically to that of Figure \ref{fig:all_opt_long}.  For each panel, the flux has been normalized at $-$250 \kms\ and the star's rest velocity (12.3 \kms) has been subtracted out.  Each feature is labeled in the inset.  Note the P Cygni profiles of the Na I D lines in the 2009 Mar and May data and that the K I line goes into emission in the three 2009 Jan epochs.}
\label{fig:rv_plot}
\end{figure*}

We have also examined the radial velocity profiles of several FELs, Balmer lines, and absorption features of our MagE spectra (see Figure \ref{fig:rv_plot}).  These profiles have been corrected for the star's rest velocity (12.3 \kms) and normalized to the continuum at $-$250 \kms.  Of note, the K I $\lambda$7699 line goes from absorption into emission for the three 2009 Jan epochs.  A very weak P Cygni profile is seen in the profile of H$\alpha$ at the 2009 Mar epoch.  While all seven epochs of data show Na I D $\lambda$5890, 5896 in emission, the last two epochs of data, 2009 Mar 5 and 2009 May 31 UT, have sharp P Cygni profiles in this feature as well, indicating that we are seeing outflowing winds along the line of sight.

\subsection{NIR Spectroscopy}

\subsubsection{Spectral Morphology and Reddening}

\begin{figure*}[htbp]
\centering
\includegraphics[scale=0.45]{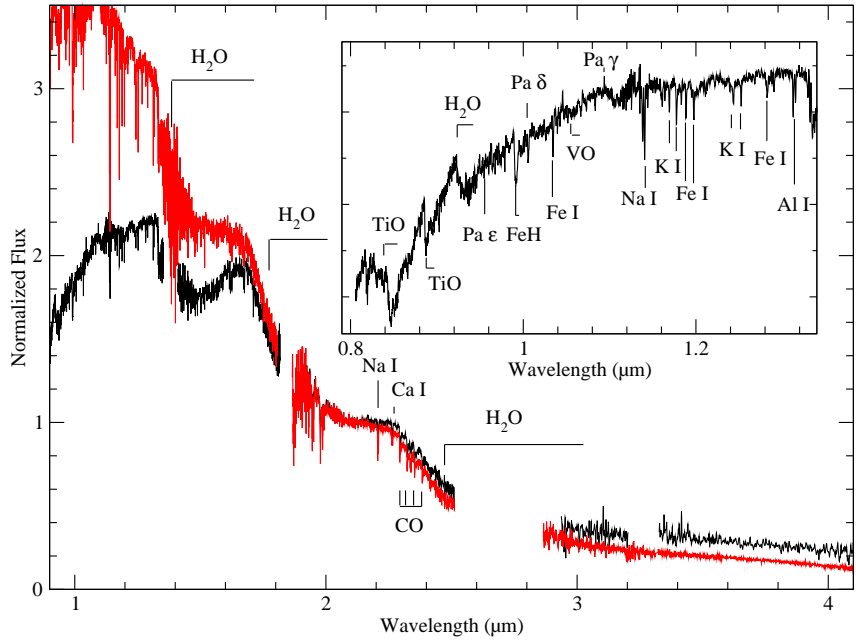}
\caption{The SpeX SXD spectra of TWA 30 at the stated epochs.  All spectra have been normalized at 2.15 $\mu$m.  The spectra have not been de-reddened.  The spectrophotometry and colors calculated from these data are listed in Table \ref{table:specphot}.}
\label{fig:all_nir}
\end{figure*}

TWA 30 was observed over several epochs spanning nearly seven months using 0.9--2.4 $\mu$m spectroscopy.  
As we did not have a comprehensive set of young (low-gravity) early-to-late M templates in the NIR, we have used field templates and carried out the same spectral typing and $A_v$ determination as before in the optical data analysis.
In Figure \ref{fig:all_nir}, we show our eight epochs of NIR data normalized in the $K$-band, which are not taken concurrently with our optical data.  From these spectra we have determined spectral types and $A_v$ measurements for all eight epochs (see Table \ref{table:specphot}) by dereddening our spectra to the following field templates\footnote{All template spectra were obtained from http://irtfweb.ifa.hawaii.edu/$\sim$spex/IRTF\_Spectral\_Library/index.html.  These spectra were taken with the same set-up but using the 0$\farcs$3 slit, instead of the 0$\farcs$5 slit we employed for our observations.} -- Gl 213 (M4 V), Gl 51 (M5 V), Gl 406 (M6 V), vB 8 (M7 V), vB 10 (M8 V).  These spectra show a large variability in the amount of reddening present ($A_v$=1.5--9.0), which is also correlated with the dereddened spectral types.  The epoch with the least amount of reddening, 2009 Feb 2, has the earliest spectral type: M4.5, while the epoch with the highest amount of reddening, 2009 May 20, has the latest spectral type: M6.  We outline a possible cause for the changing spectral types in both the optical (M4--M5.25) and NIR (M4.5--M6) in $\S$4.2.  Given the variability in spectral type in both the optical and NIR regimes, we adopt a spectral type of M5~$\pm$~1.

We have combined the first of these epochs, 2008 Dec 4 UT, with our only longer wavelength (1.9--4.20 $\mu$m) spectroscopic data from 2008 Dec 5 UT.  A significant amount of reddening is observed in some of our spectra on day-timescales; however, the $K$-band ($\sim$1.9--2.4 $\mu$m) overlap between these two datasets is virtually identical.  We compare this wider spectral energy distribution to the M5 field template, Gl 51, in Figure \ref{fig:near_IR_spec}.  Both spectra have been normalized in $K$-band (at 2.1 $\mu$m), showing that TWA 30 is redder than the M5 template, although the longer wavelength $L$-band is less affected and matches the template reasonably well.   We have chosen an M5 for comparison, as when we dereddened this spectrum the M5 template provided the best fit in spectral slope.  

We have also measured spectrophotometry on these data to determine 2MASS $JHK_s$ colors, as well as on our two lower resolution NIR spectra from 2009 Jun 28 and 29 UT, and report these in Table \ref{table:specphot} along with the 2MASS PSC measurements from 1999 Mar 24 UT.   The epoch of our data with the least amount of reddening, 2009 Feb 2 ($A_v$=1.5), has $JHK_s$ magnitudes very similar to those from the 2MASS PSC ($\Delta$~=~0.01--0.15 mag) epoch.  While these spectra have been flux calibrated with nearby A0 V stars, they have not been absolutely flux calibrated with photometry, and therefore do not take into account slit losses.  \cite{2009ApJS..185..289R} estimate that in photometric conditions absolute spectrophotometry performed on SpeX spectra calibrated with nearby A0 V stars are accurate to 10\%, while relative spectrophotometry are accurate to a few percent.
 
\begin{figure*}[htbp]
\centering
\includegraphics[scale=0.6]{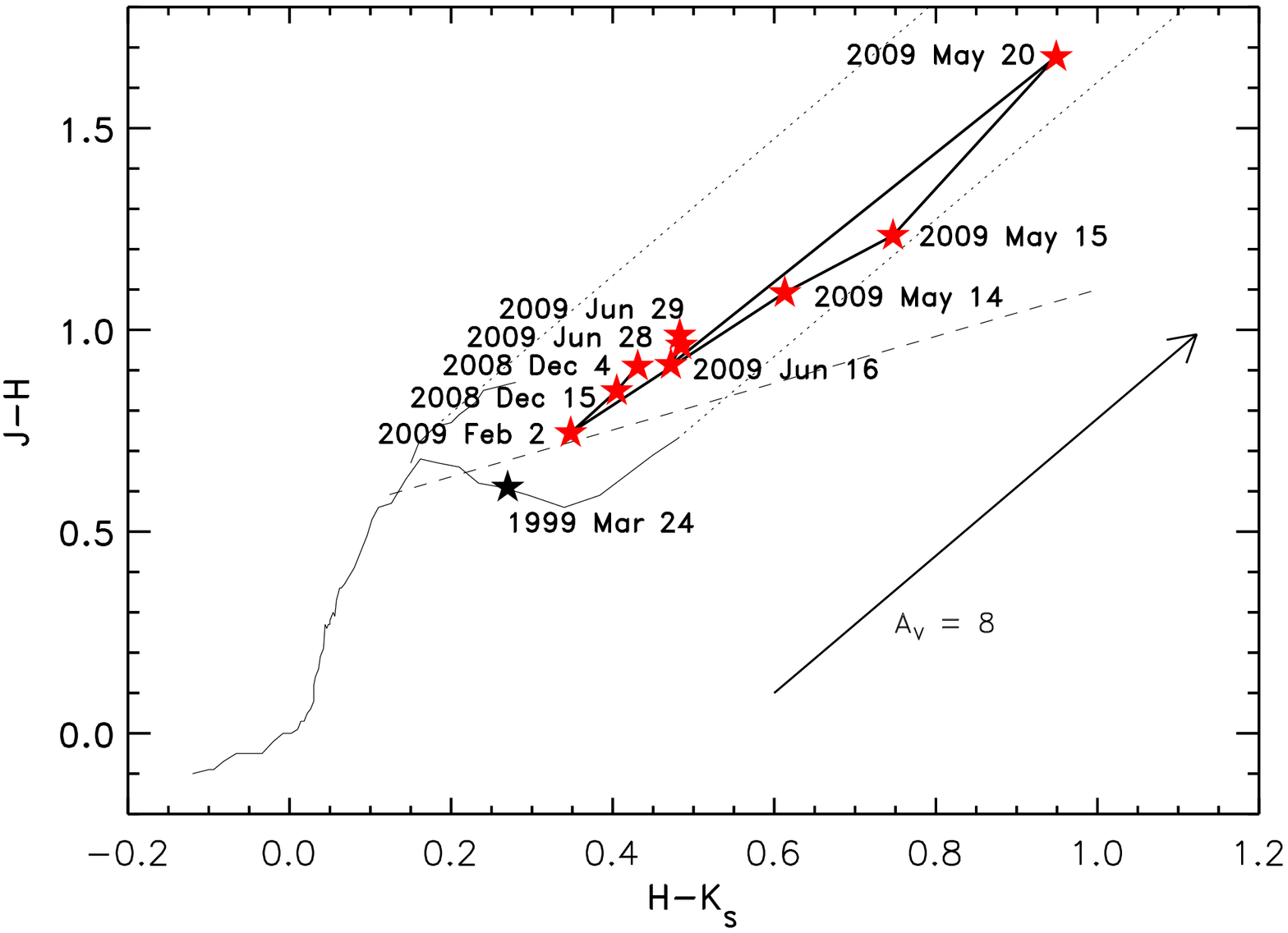}
\caption{The NIR colors of TWA 30 at each epoch of our SpeX data ({\it red stars}; see Table \ref{table:specphot}) and one epoch ({\it black star}; 1999 Mar 24 UT) from the 2MASS PSC.  The dwarf track ($\ge$M0 from \citealt{1992ApJS...82..351L}; earlier types from \citealt{1995ApJS..101..117K}) is the lower solid line and the giant track \citep{1988PASP..100.1134B} is the upper solid line, the vectors for reddened dwarfs are shown as two dotted lines, the cTTS locus \citep{1997AJ....114..288M} is shown as a single dashed line, and the arrow at the lower right indicates a reddening of A$_{v}$=8.  Both the dwarf and giant tracks have been transformed to the CIT photometric system, which is similar to the 2MASS system \citep{2001AJ....121.2851C}.  The red stars are connected by a solid straight line to delineate the sequence of observations.}
\label{fig:nirplot}
\end{figure*}

The colors calculated from these measurements appear to be accurate between our SXD and prism spectra (Table \ref{table:specphot}), with identical $H-K_s$ colors and similar $J-H$ ($\Delta$~=~0.06 mag) colors.   These values are shown in $JHK_s$ color space in Figure \ref{fig:nirplot}.  Only the 2MASS PSC colors of TWA 30 lie along the dwarf track.  All other epochs lie along the reddening vector described by \cite{1989ApJ...345..245C}.  The sparse coverage of our data shows a full cycle over a six month period, however, the rapid increase in $A_v$ over the period 2009 May 14--20 may indicate that the cycles are much shorter if this behavior is periodic.

\subsubsection{Emission and Absorption Lines}

\begin{figure*}[htbp]
\centering
\includegraphics[scale=0.6]{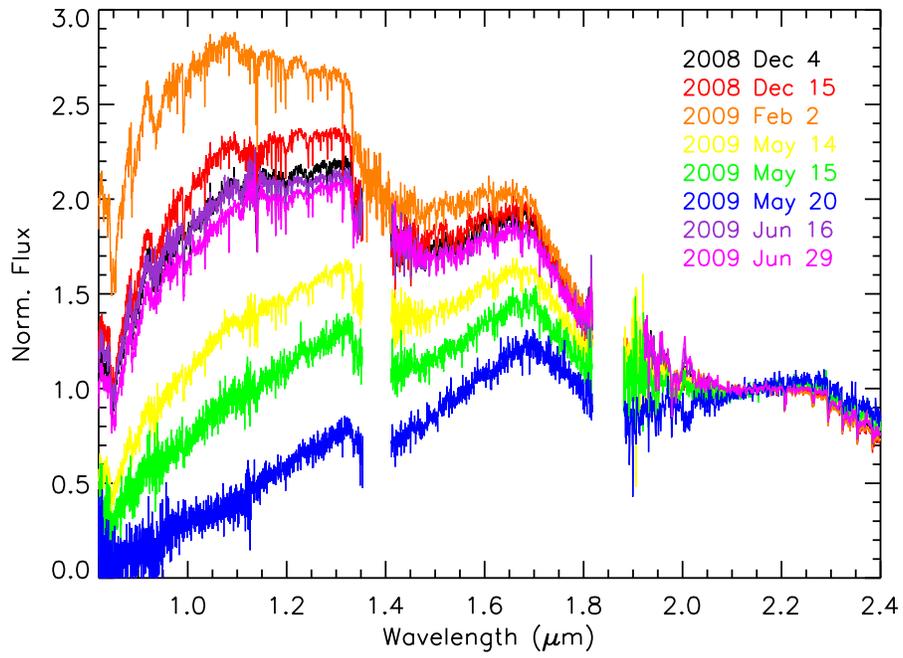}
\caption{The combined SXD and LXD SpeX spectra of TWA 30 (black) obtained on 2008 Dec 4--5 UT shown in comparison to the M5 V field template Gl 51 (red).  The spectrum of TWA 30 has not been de-reddened here or in the inset.  Both spectra have been normalized at 2.1 $\mu$m.}
\label{fig:near_IR_spec}
\end{figure*}

\begin{figure*}[htbp]
\centering
\includegraphics[scale=0.6]{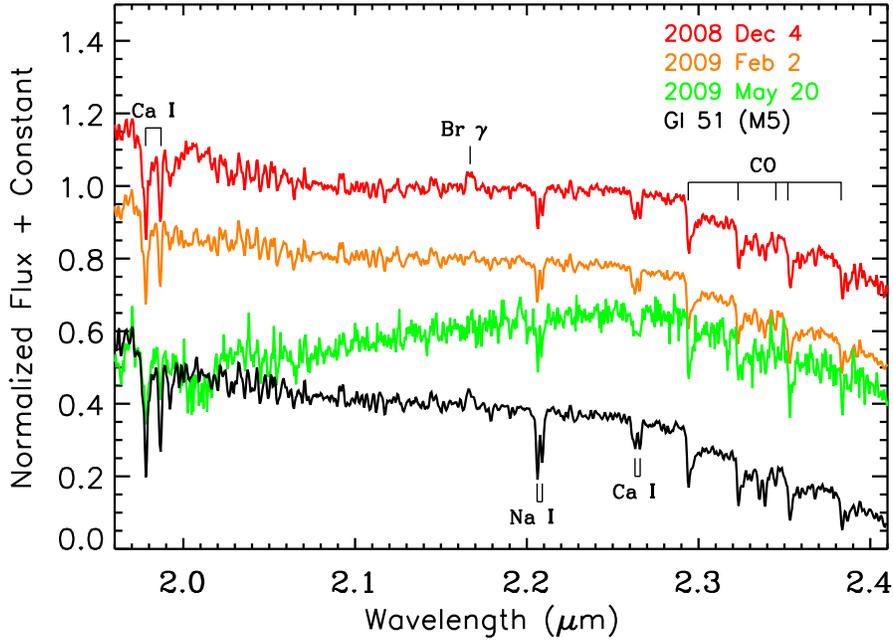}
\caption{SpeX SXD spectra in the $K$-band region of TWA 30 at the three stated epochs (red, orange, green) shown in comparison to the M5 V field template Gl 51 (black).  All spectra have been normalized at 2.15 $\mu$m and are separated by constants of 0.2 along the y-axis.  Note that the alkali doublet Na I is much stronger in the M5 template than in the spectra of TWA 30.  Br $\gamma$ is weakly present in the 2008 Dec 4 UT spectrum and in the Gl 51 template but is absent in the 2009 Feb 2 and 2009 May 20 data.}
\label{fig:Kband}
\end{figure*}

As seen in the optical data, typical surface gravity indicators such as the alkali metal lines, K I and Na I, are weaker in TWA 30 than in the field M5 template (see Figures \ref{fig:near_IR_spec} and \ref{fig:Kband}).  While H Balmer emission in the optical spectrum of TWA 30 is prevalent, H Paschen (Pa) is in {\it absorption} in the NIR, with the exception of Pa $\gamma$, which is seen in emission in our lower resolution SXD data (see Figure \ref{fig:near_IR_spec}).  A typical indicator of on-going accretion, H Brackett $\gamma$ can be seen only weakly in emission at 2.166 $\mu$m in the $K$-band spectrum of the 2008 Dec 4 UT data and is absent in the 2008 Dec 15 and 2009 Feb 2 UT data (see Figure \ref{fig:Kband}).

\subsection{Estimated Mass and Age}

To derive the stellar mass and age we have used two estimates of the spectral type: (1) using the 2MASS colors, which imply a spectral type of $\sim$M4.5, although this probably includes accretion continuum excess emission (see $\S$4.2) and (2) using the latest NIR spectral type, M6, which likely has the least accretion continuum excess (see $\S$4.2), although the spectral type may be slightly too late due to our use of field templates for the NIR data.  

For the first estimate, we assume that the 2MASS data point (K$_s$=8.77~$\pm$~0.02 mag) has A$_V$, A$_K$~=~0.  From the T$_{eff}$ scale\footnote{\cite{2008ApJ...684..654L} estimate an uncertainty of $\sim$100 K in this relation.} for cTTS of \cite{2003ApJ...593.1093L}, a spectral type of M4.5 gives an effective temperature of 3200~$\pm$~100 K.  The bolometric correction in $K$-band (BC$_{K}$) from \cite{2004AJ....127.3516G} begins at spectral type M6, yielding BC$_K$~=~3.03~$\pm$~0.13 mag for M6 and BC$_K$~=~3.06~$\pm$~0.13 mag for M7.  Although M4.5 is earlier than the beginning of this relation, the change in BC$_K$ over 1 spectral type ($\Delta$~=~0.03 mag) is small in comparison to the root-mean-square (RMS) in the uncertainty (0.13 mag).  We therefore adopt a value of BC$_K$~=~3.0~$\pm$~0.1 mag on the CIT photometric system.  Applying the transformation of \cite{2001AJ....121.2851C}, 2MASS $K_s$= CIT $K$ $-$ 0.024 mag, with negligible uncertainties and a color term of zero.  Hence, 2MASS $K_s$~=~8.77~$\pm$~0.02 mag transforms into CIT $K$~=~8.79~$\pm$~0.02 mag.  For a distance of 42~$\pm$~2 pc (the kinematic distance), the distance modulus is 3.12~$\pm$~0.10 mag, yielding CIT M$_K$~=~5.67~$\pm$~0.10 mag.  Applying the bolometric correction yields M$_{bol}$~=~8.7~$\pm$~0.1 mag.  Adopting the IAU standard of M$_{bol,\sun}$~=~4.75 mag, log L$_{bol}$/L$_{bol,\sun}$~=~$-$1.58~$\pm$~0.04 dex.  With this T$_{eff}$ and L$_{bol}$, the evolutionary models of \cite{1998A&A...337..403B} give an age of 10 Myr and a mass of 0.16 M$_{\sun}$.

For the second estimate, we use the latest NIR spectral type derived, M6, in 2009 May 14 and 20 with the corresponding 2MASS $K_s$ and A$_V$ estimates listed in Table \ref{table:specphot}.  Using A$_K$/A$_V$=0.112 \citep{1985ApJ...288..618R} and the same CIT $K$ transformation as listed above to correct the NIR magnitudes of these two epochs, we find CIT $K$ (A$_K$=0)~=~9.05~$\pm$~0.04 mag.  With the same distance modulus as above, CIT M$_K$~=~5.93~$\pm$~0.11 mag.  The BC$_{K}$ correction from \cite{2004AJ....127.3516G} for an M6 is 3.03~$\pm$~0.13 mag, yielding M$_{bol}$~=~9.0~$\pm$~0.2 mag and log L$_{bol}$/L$_{bol,\sun}$~=~$-$1.70~$\pm$~0.08 dex.  The T$_{eff}$ scale of \cite{2003ApJ...593.1093L} for an M6 gives $T_{eff}$~=~3000~$\pm$~100 K.  With this T$_{eff}$ and L$_{bol}$, the evolutionary models of \cite{1998A&A...337..403B} give an age of 4 Myr and a mass of 0.08 M$_{\sun}$.

As these two spectral types and NIR photometric measurements cover nearly the full range of our measurements, the age of TWA 30 likely lies somewhere between these extremes.  The mass inferred from the optical data is likely an overestimation, due to the inclusion of accretion luminosity, while the age inferred from the NIR data is likely an underestimation due to our use of field templates instead of young templates to infer the NIR spectral type.  Given these uncertainties, we adopt an age of 7~$\pm$~3 Myr and a mass of 0.12~$\pm$~0.04 M$_{\sun}$.  This age estimate agrees with the estimated age of the TWA ($\sim$8 Myr), and the mass estimate places it near the substellar boundary at $\sim$0.08 M$_{\sun}$ \citep{2000ARA&A..38..337C}.

\subsection{X-ray Activity}

The RASS Faint Source Catalog lists an X-ray detection of (2.5~$\pm$~1.0) $\times$ 10$^{-2}$ counts s$^{-1}$ at a distance of 17$\farcs$6 from the 2MASS epoch position of TWA 30.  The hardness ratio, HR1, is (A$-$B)/(A+B) where band-A covers 0.1--0.4 keV (soft X-rays) and band-B covers 0.5--2.0 keV (hard X-rays).  The conversion factor, CF, from counts s$^{-1}$ to X-ray flux (\ergcms) for RASS data of late-type stars is derived by \cite{1995ApJ...450..401F} as CF~=~(8.31 + 5.30$\times$HR1) $\times$ 10$^{-12}$ \ergcms.  The CF for TWA 30 is thus 7.25 $\times$ 10$^{-12}$ \ergcms\ and when combined with the RASS count rate, yields an X-ray flux of F$_X$~=~1.8 $\times$ 10$^{-13}$ \ergcms.  With an adopted distance of 42 pc, we then derive log L$_X$~=~28.6 dex.
From the average bolometric luminosity calculated above (log L$_{bol}$/L$_{bol,\sun}$~=~$-$1.64~$\pm$~0.09 dex; L$_{bol}$~=~8.8 $\times$ 10$^{31}$ \ergs, assuming L$_{bol,\sun}$~=~3.85 $\times$ 10$^{33}$ \ergs), this yields log L$_X$/L$_{bol}$~=~$-$3.34 dex.  For early to mid-M dwarfs, log L$_X$/L$_{bol}$ saturates at $\sim-$3 dex \citep{1998A&A...331..581D}.  
For T Tauri stars, both accreting and non-accreting, down into the BD domain, log L$_X$/L$_{bol}$ ranges from $\sim-$3 to $-$4 dex (e.g., the Taurus XMM survey; \citealt{2007A&A...468..353G}), which the value of TWA 30 falls into, i.e., is typical.

\section{Discussion} 

\subsection{The Disk Fraction of the TWA} 

Our observations provide clear evidence for the presence of an actively accreting disk and outflow around TWA 30; however, it is not the only disk-bearing low-mass source in the TWA. \cite{2005ApJ...631.1170L} measured a 25\% disk fraction from a survey of 24 candidate TWA members observed from {\it Spitzer Space Telescope} mid-IR measurements, including two M1 dwarfs (TWA 7 and TWA 13, the latter of which has been discounted by \citealt{2009ApJ...698.1068P}).  Other M dwarf members -- TWA 14 (M0.5) and TWA 5A (M2.5) -- have indications of passive circumstellar disks such as broad asymmetric Balmer emission \citep{2001ASPC..244..245M,2003ApJ...593L.109M} but lack NIR excess emission (however, \citealt{1999ApJ...521L.129J} find a modest 10 $\micron$ excess emission for TWA 5A).   Two of three substellar systems observed in the mid-IR -- TWA 28 and TWA 27A (2M1207A), and possibly TWA 27B  \citep{2003ApJ...593L.109M,2004A&A...427..245S,2006ApJ...639L..79R,2008ApJ...676L.143M,2008ApJ...681.1584R,2009ApJ...696.1589H} -- also exhibit signatures of IR excess, accretion and outflow, indicating that the same processes occur in low-mass stars and brown dwarfs \citep{2007ApJ...659L..45W,2009ApJ...696.1589H}. 

\begin{figure*}[htbp]
\centering
\includegraphics[scale=0.7]{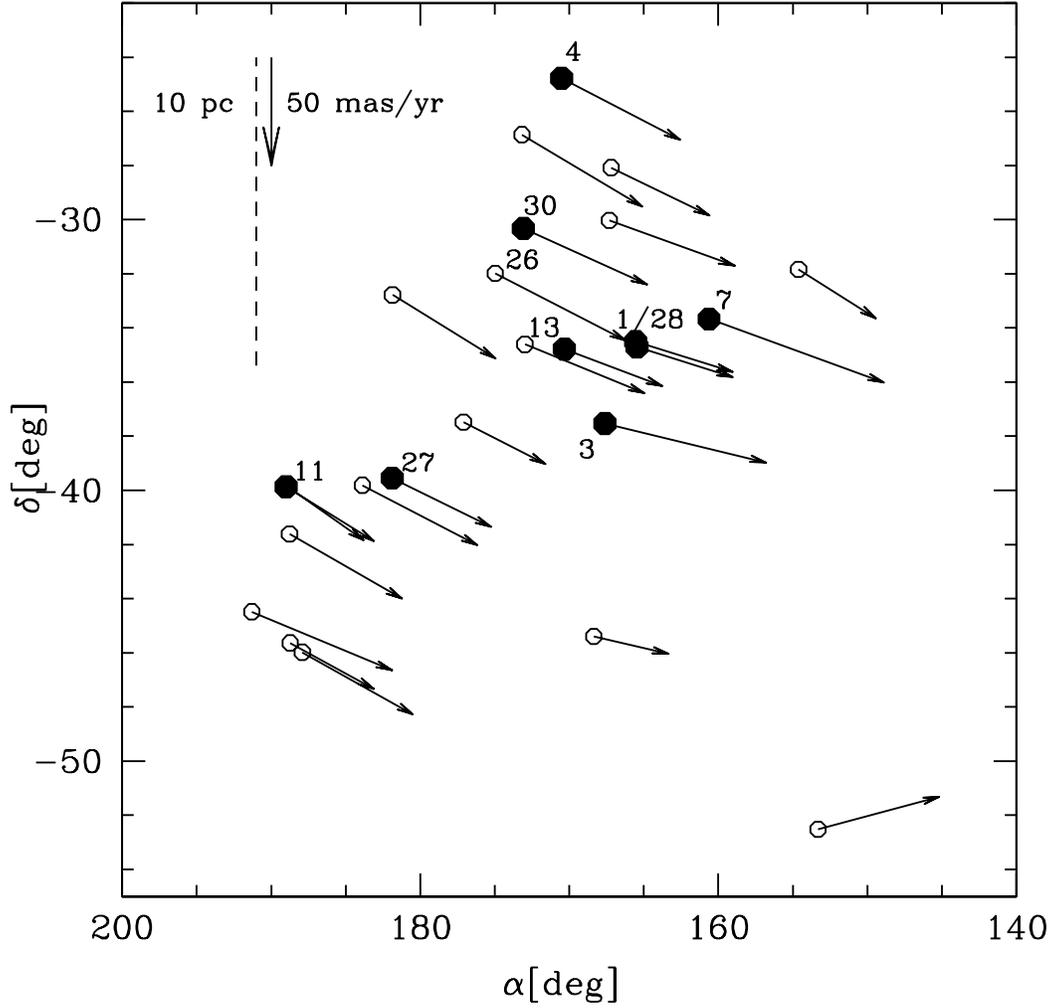}
\caption{The spatial locations of all TWA members with disks (filled circles) and without disks (empty circles) based on mid-IR excesses, with the exception of TWA 30, which has no mid-IR data available but shows spectroscopic signs of a disk.  Members with a disk are also labeled by their TWA designation.  Note that most members with a disk reside in the north-west quadrant of the TWA.  The short-dashed line marks a distance of 10 pc at the mean distance of the TWA members (53~$\pm$~3 pc), and the arrow indicates the scale length for the proper motions of members.}
\label{fig:pm}
\end{figure*}

With the addition of TWA 30 as a disk-bearing low-mass member of the TWA, we updated the disk fraction of the TWA considering the following systems listed in \cite{2005ApJ...634.1385M} as {\it bona fide} members: TWA 1--11, 13--16, 20, 21, 23, 25--28 (see \citealt{2008hsf2.book..757T} and \citealt{2005MNRAS.357.1399L} for slightly different membership lists).  While we consider TWA 29 (DENIS J124514.1$-$442907) to be a potential member, no mid-IR observations have yet been reported so we do not include it in our census.  Of these systems, many of which are known binaries (e.g., \citealt{2003AJ....126.2009B}), mid-IR observations (listed above) have confirmed circumstellar disks around TWA 1, 3AB, 4AB, 7, 11A, 27A(B), and 28 (shown in RA/decl space in Figure \ref{fig:pm}).  Including TWA 30 increases the disk-bearing population to 8 of 23 systems, or 35$^{+11}_{-8}$\%.   This is a 40\% increase (at $\gtrsim$1$\sigma$) to the estimate provided by \cite{2005ApJ...631.1170L}.

\subsection{An Inclined Stellar Axis to an Edge-On Disk?}

There are seven striking features of the TWA 30 system that any scenario
must explain:

\begin{enumerate}
\item The presence of a large number of FELs indicative of jets, with line-center velocities both blue- and red-shifted with
respect to the stellar radial velocity but by very small amounts (generally $\lesssim$10 \kms);\\
\item Temporal variations in the NIR colors and spectrum, with increasing reddening associated with decreasing NIR flux;\\
\item Temporal variations in the reddening of the optical spectrum as well;\\
\item Temporal variations in the underlying optical and NIR spectral types inferred after reddening effects are removed, with the earliest spectral type (optical: M4, NIR: M4.5 -- non-contemporaneous) associated with the lowest reddening and with the latest type (optical: M5.25, NIR: M6 -- non-contemporaneous) associated with the highest reddening;\\
\item Changes in the equivalent widths of the optical photospheric absorption lines of Li I and K I, with the least absorption when the optical spectral type is earliest; \\
\item The sporadic appearance of P Cygni profiles in the Na I D resonance doublet, and more weakly, in H$\alpha$, with the P Cygni profiles being weakest/absent when the absorption in the Li I and K I lines mentioned above is least; and\\
\item Variations in the equivalent widths of the FELs, with the smallest widths appearing in the last epochs of optical data when the P Cygni profiles are strongest.\\
\end{enumerate}

\subsubsection{Evidence from FELs for an Edge-On Disk} 

In regards to the first point, jets from accreting cTTS reach velocities\footnote{The jet velocity is expected to scale roughly as the escape velocity from the object \citep{2004ApJ...615..850M}, which is very similar for a 0.5 $M_{\odot}$ fiducial cTTS and the $\sim$ 0.12 $M_{\odot}$ TWA 30.} of order 200--300 \kms.  If the FELs in TWA 30 are assumed to trace such a jet, as in other cTTS, then their very low apparent velocities imply a jet that is nearly aligned with the plane of the sky.  This is supported by the very low velocities seen even in the [NII] 6583~{\AA} line, which invariably exhibits only a high velocity component in cTTS \citep{1997A&AS..126..437H}.  By extension, the accretion disk, perpendicular to the jet, must be seen very close to edge-on.  This conclusion is further supported by the presence of both blue- and red-shifted FELs; significant deviations from an edge-on viewing angle in cTTS produce only blue-shifted FELs, as the disk obscures the receding red-shifted jet lobe (e.g., \citealt{2009ApJ...691L.106W}).  Further evidence for an edge-on geometry from other signatures is discussed further below.

\subsubsection{Evidence for Stellar Reddening due to Disk Dust}

The occasional yet significant reddening of the NIR spectrum (point 2), accompanied by a simultaneous dimming of the star in NIR bands, is most easily explained as stellar reddening and occultation by disk dust within the line of sight.  In the edge-on case, this dust may be entrained with the gas at the base of the outflow / accretion funnel-flow near the inner edge-of the disk, or signify a disk warp produced in this region by the interaction of a tilted stellar magnetosphere with the disk inner edge, as proposed for the cTTS AA Tau \citep{1999A&A...349..619B,2003A&A...409..169B,2007A&A...463.1017B}.  

Similarly, the variable reddening observed in the optical spectrum (point 3) may then also be interpreted as extinction/reddening caused by disk dust in the line of sight.  Unfortunately, we do not have simultaneous optical photometry to verify this, as possible in the NIR.  Nevertheless, the similar relationship between reddening and spectral type in the optical and NIR supports this conclusion, as described further below.

\subsubsection{Evidence for Accretion-related Veiling and Variability}

In both the optical and NIR, the underlying spectral type varies by similar amounts after the reddening effects are accounted for (point 4): from M4 to M5.25 in the optical (Table \ref{table:AVs} and Figures \ref{fig:mage_spec1}, \ref{fig:mage_spec2}, and \ref{fig:mage_spec3}), and from M4.5 to M6 in the NIR (Table \ref{table:AVs}).  For an accreting object, such changes are most easily ascribed to changes in veiling, i.e., changes in the excess continuum emission due to accretion.  In this view, the earliest spectral type corresponds to maximum accretion, and the latest spectral type to minimum accretion.  

This interpretation is supported by the behavior of the photospheric lines of Li I $\lambda$6708 and K I $\lambda$7699 (point 5).  
Both show variable absorption: the smallest absorption equivalent widths (indeed, K I goes into emission) are seen in 2009, when the optical spectral type is earliest, while the largest absorption widths occur in 2009 Mar and May, when the optical types are the latest.  For photospheric absorption, this cannot be due simply to changes in the stellar continuum (e.g., due to suppression of the continuum by dust occlusion), since the lines and continuum would be equally affected without changing the equivalent width.  Instead, it points to a significant enhancement of accretion during this period which partially fills in (veils) the Li I absorption through excess continuum emission and also produces excess line emission in K I.  The observed correlation between decreasing absorption strength (i.e., enhanced accretion) and earlier spectral type then confirms that the changes in the underlying optical spectral type are due to veiling effects.  By Occam's razor, the similar variations in the NIR spectral type may then also be attributed to accretion-induced veiling effects.  Note that if this veiling interpretation is correct, then the true spectral type of the star is likely to be $\sim$M5.25--M6, i.e., the latest and thus least veiling-affected types we see in the optical and NIR.  Nevertheless, given the additional complexity introduced by reddening variability (see next sub-section), we have simply adopted M5~$\pm$~1 which covers the full range seen in the data; a more accurate classification will have to await time-resolved high-resolution spectroscopy.

\subsubsection{Observed Relationship Between Reddening, Accretion and Wind Signatures}

Moreover, we also see the least reddening $A_V$ in both the optical and the NIR, when the underlying spectral type is earliest, and intermediate (in the optical) to highest (in the NIR) $A_V$ when the inferred spectral type is the latest (point 4).  If the spectral type changes are due to variable accretion and if the extinction is due to dust in the line of sight, as we argue, then this implies that the dust is mostly anti-correlated with accretion.  A physical interpretation of this is supplied further below.    

P Cygni profiles (point 6) are clear signatures of cool outflowing winds along the line of sight moving towards the observer (e.g., \citealt{2003ApJ...599L..41E} and references therein).  Their sporadic appearance in the Na I D lines, and once weakly in H$\alpha$ (2009 Mar, when the Na I D P Cygni profiles are also strongest), implies that the wind from TWA 30 intersects our line of sight only occasionally.  Moreover, the P Cygni profiles are strongest in 2009 Mar and May, when the apparent optical spectral type is latest, M5.25, and absorption in photospheric Li I and K I is greatest, i.e., by our above arguments, when the accretion is weakest.  Conversely, the P Cygni profiles are weakest/absent in 2009 Jan, when the optical spectral type is earliest, M4, and the Li I and K I absorption is weakest -- i.e, when accretion is strongest.  It is noteworthy that the Na I D doublet evinces the strongest, somewhat red-shifted emission in the same 2009 Jan spectra.  This cannot be explained simply by the absence of {\it blue}-shifted P Cygni emission, but instead again indicates enhanced accretion when the P Cygni wind signatures are weakest.

There are two possible explanations for this behavior.  The first is that the P Cygni profiles are always present at similar strength, but completely filled in by accretion-induced veiling when accretion is strongest.  The other is that there is a real weakening of the wind signatures when accretion is strongest.  While some veiling of absorption profiles clearly occurs during enhanced accretion (e.g., in Li I), we suggest that the second effect -- a real weakening of the wind -- is likely to be present as well, due to the following physically-motivated argument.         

\subsubsection{A Physical Interpretation: An Inclined Stellar Magnetosphere Interacting with the Disk Inner-Edge}

We have noted earlier that the extinction/reddening signatures, caused by dust in the line of sight, are anti-correlated with accretion, while now we find that the P Cygni wind signatures show the same behavior.  Both can be explained simultaneously by the interaction of a tilted stellar magnetosphere interacting with the inner edge of the accretion disk.  In this case, accretion is energetically preferred along one half of the inner edge circumference (where the field lines bend inwards towards the star) while outflow is preferred along the other half (where the field lines bend outwards towards the disk).  As the star rotates, an observer viewing the disk edge-on along either the top or bottom disk surface would preferentially see outflow and accretion 180\degree\ out of phase in time.  This is precisely what is seen, for instance, in the cTTS SU Aur (e.g., \citealt{1995ApJ...449..341J}), and would explain the alternating P Cygni wind signatures and enhanced accretion we observe in TWA 30.  Furthermore, the interaction of an inclined magnetosphere is expected to create a disk warp on theoretical grounds \citep{2000A&A...360.1031T}, which would explain the variable reddening
observed in TWA 30.  In the case of AA Tau, where such a warp has also been invoked \citep{2003A&A...409..169B}, it appears associated with the accretion funnel flow, i.e., is correlated instead of anti-correlated with the accretion as we observed for TWA 30.  However, \cite{2000A&A...360.1031T} show that the specific sense of the warp -- i.e., whether the warp vertical extension is parallel or anti-parallel to that of the accretion funnel flow or wind -- depends on particular assumptions about the nature of the field at the disk inner edge.  Hence, a disk warp associated with the wind base is possible as well.  In summary, this single physical situation can explain the variable extinction, variable wind signatures, variable accretion, and apparent anti-correlation of the two former with the latter.  On the other hand, if only increased veiling were responsible for the disappearance of the wind-related P Cygni profiles during periods of enhanced accretion, we have no obvious reason for the reddening to be anti-correlated with accretion.   We therefore propose that there is in fact a real change in the wind signatures.  This must be checked through future high-resolution, time-resolved optical spectroscopy, wherein the wavelength-dependent veiling can be characterized in detail.    

\subsubsection{Further Evidence for an Edge-On Geometry, and Implications for Accreting Gas}

We further see that: the P Cygni profiles are strong in Na I D but weak in H$\alpha$; the Na I D absorption is only moderately blue-shifted from the stellar rest-frame (mean $\sim$25 \kms; Figure \ref{fig:rv_plot}); and the H$\alpha$ emission varies by a factor of 1.8 (following the same trend as other accretion signatures -- such as veiling and line emission in K I and Na I D --  of being enhanced in 2009 Jan compared to 2009 Mar and May).  All these are consistent with the edge-on hypothesis.  In this geometry, our line of sight intersects the cool dense base of the flow where the resonance lines of Na I and K I are easily excited but not H$\alpha$, so we are more sensitive to changes in accretion and outflow in the former lines than in the latter; this is also where a magneto-centrifugal wind has the smallest velocity (and we see an even smaller component of it in our edge-on line of sight; \citealt{2007IAUS..243..171E,2007ApJ...657..897K}), yielding only a small shift in radial velocity in wind signatures relative to the stellar rest frame. 

In this regard, we also note that in many edge-on systems red-shifted, {\it inverse} P Cygni absorption is seen superimposed on the H$\alpha$ emission, arising from infalling gas seen against the hotter accretion shock on the stellar surface.  No such feature is apparent in our H$\alpha$ profiles.  This may be because the accreting gas is optically thick, completely obscuring the shock region, as supported by the lack of detectable emission in He I at 4471, 6678, or 7065 {\AA}, as well as very weak narrow emission in He I 5876 {\AA} (the strongest line in the series).  In cTTS, He I appears to be produced in both the accretion flow, giving rise to a broad component, and in the shock, yielding a narrow component.  The absence of a broad He I component in our data is consistent with an edge-on geometry, since this is a high-temperature line that would not be excited at the cool base of the flow.  The absence or extreme weakness of a He I narrow component in the presence of other accretion signatures, further suggests that the shock region is obscured by optically thick accreting gas.

\subsubsection{Evidence for Time-Variability in the FELs, and a Physical Interpretation}

The scenario sketched so far is consistent with spatial variations alone: the changes in the accretion/outflow/extinction signatures, due to inner-disk interactions with a tilted stellar magnetosphere, occur as the accretion flow, outflow and disk warp rotate into and out of our line of sight.  As such, these signatures should be periodic on the scale of the rotation period of the disk inner edge.  In cTTS, this is usually on the order of days.  Our observations are very poorly sampled at such short periods, and thus remain consistent with this picture of spatial changes alone.  We note that the very strong increase in NIR reddening in 2009 May over a period of only 6 days hints at changes on the timescales expected from rotational effects.  This will need to be verified by future finely time-sampled monitoring.  

However, spatial variability alone cannot be the whole story.  Considering the FELs in TWA 30 (point 7), we note first that the equivalent widths of nearly all of these are weakest in the 2009 Mar and May optical data and strongest in the 2009 Jan optical data.  This could be explained if the stellar photosphere were severely occulted in 2009 Jan relative to 2009 Mar and May; i.e., if the FELs were seen in 2009 Jan against a much dimmer continuum.  However, this appears in contradiction to our argument above, where the optical data indicate that it is 2009 Jan when the $A_V$ is least (indicating minimum dust reddening and occlusion) and 2009 March and May when the $A_V$ is intermediate to highest (indicating maximum reddening/occlusion effects).  Moreover, the equivalent widths of different FELs change by different amounts, while they should all change by the same proportion if it is the stellar continuum that is being suppressed.  The FELs are also expected to arise on scales of at least a few AU from the star, and thus cannot themselves be expected to be affected by any occultation mechanisms operating near the star (e.g., a disk warp).

The only alternative is that the FELs are intrinsically time-variant. FEL time variations may arise as the jet encounters inhomogeneities in its path, or represent temporal changes in the jet launching.  If the latter occurs in TWA 30, then the strengthening of the P Cygni profiles in 2009 Mar and May may indicate an actual increase in the outflow rate at these epochs instead of just the outflow rotating into our line of sight.  Such a scenario is not inconsistent with the FELs being weakest in these epochs.  An FEL can only arise when the critical density of that particular species is reached, at a distance of tens of AUs for cTTS to perhaps a few AU for brown dwarfs \citep{2009ApJ...691L.106W}.  At the very least, there must be a time-lag between the observation (via P Cygni profiles) of a strengthening in the wind at its base, and the corresponding increase in the FELs a few AU away, representing the jet travel time.  Assuming a fiducial jet velocity of $\sim$300 \kms\ and a distance of $\sim$5 AU to the critical density region, yields a time lag of $\sim$30 days.

Our interpretation of the FEL variations is then simply that at some point in the past there was a strengthening in the outflow, which led (after some time-lag) to the brightening of the FELs in the 2009 Jan spectra.  By the time of the 2009 Mar and May spectroscopy, the FELs have faded back to their normal levels.  The fact that these epochs coincide with an increase in accretion and wind signatures near the star, respectively, would simply be coincidental.  This is a plausible scenario if changes in the latter signatures occur episodically (and maybe periodically) on timescales of days.  We note further that, relative to the 2008 Nov data, the FELs in the 2009 Jan data are stronger: in [O I] $\lambda$6300 by 49\%, in [S II] $\lambda$6731 by 165\%, and in [N II] $\lambda$6583 by 63\% (see Table \ref{table:AVs}).  This may be understood by noting that the requisite critical electron density is highest for [O I], smaller for [S II] and least for [N II].  Thus, for a jet whose streamlines diverge (i.e., density decreases) with increasing distance, the [O I] emission region lies closest to the star, [S II] appears further out, and [N II] lies furthest away, as found earlier by \cite{1997A&AS..126..437H} for cTTS.  We thus posit that the past outflow enhancement has nearly reached the [O I] critical region by 2008 Nov but has yet to arrive at the [S II] and [N II] locations; by 2009 Jan the [O I] FELs have plateaued and the outflow enhancement has arrived in the [S II] region, brightening this line considerably, but has not quite reached [N II] yet; and at last, by 2009 Mar and May, the enhancement has passed all these regions, and all the FELs have faded away again.

To clarify this model, multiple observations are needed: finely time-sampled photometry to measure the periodicity of the disk rotation/outflow, mid-IR photometry to detect the disk, mid-IR spectroscopy to detect the presence of silicates and infer grain sizes, high resolution broadband imaging to resolve the disk, high resolution narrowband imaging to resolve the outflows, polarimetry to investigate the dust-disk geometry, spectroastrometry to spatially map the outflow, and sub-mm observations to measure the disk mass.   As one of the few very low-mass stars/brown dwarfs for which many of these observations are possible, TWA 30 could prove to be an invaluable case study for accretion and outflow processes at the low-mass end. 

\section{Conclusion}

We have identified a new and unusual member of the TW Hydrae Association, TWA 30, which has emission lines of [O I], [O II], [O III], [S II], and Mg I] near the stellar rest velocity, indicating it is powering an outflow.  The temporal changes in reddening, absorption/emission EWs, and line profiles, particularly for Na I D, suggests a stellar magnetospheric axis inclined with respect to a disk that is viewed nearly edge-on.  The presence of forbidden line emission [OIII] and Mg I] in the spectra of TWA 30 marks the first time these lines have been seen in a young star not residing in a molecular cloud and are rarely present even amongst cTTS in star formation regions.  Both the optical and NIR spectral types appear to vary in accordance with the reddening from M4--M5.25 and from M4.5--M6, respectively, so we have adopted a spectral type of M5~$\pm$~1.  We suggest that the earlier spectral types are seen at periods of high accretion and include continuum excess emission.  The spectral type of TWA 30 therefore might be as late as M5.25--M6.  From evolutionary models, we estimate an age of 7~$\pm$~3 Myr and a mass of 0.12~$\pm$~0.04 M$_{\sun}$.  The close proximity of TWA 30, 42~$\pm$~2 pc, makes it an excellent target for follow-up studies to spatially resolve the outflow and circumstellar disk.  With the inclusion of TWA 30, we have updated the disk census of the TWA, finding that 35$^{+11}_{-8}$\% of observed TWA members still retain their circumstellar disks, a higher ratio than previous estimates.

\acknowledgments
We thank our anonymous referee whose comments improved the quality of this paper.  We are grateful to George Herbig, Mike Cushing, Nathan Smith, Silvia Alencar, Silvie Cabrit, Bo Reipurth, Klaus Hodapp, Brendan Bowler, Kevin Covey, George Wallerstein, and  Suzanne Hawley for useful discussions.  We also thank our telescope operators at Magellan: Mauricio Martinez, Hern\'{a}n Nu\~{n}ez, and Ricardo Covarrubias, and at IRTF: Paul Sears, Bill Golisch, Dave Griep.  D.L.L. thanks Dave Sanders and George Herbig for financial support.  This research has benefitted from the M, L, and T dwarf compendium housed at DwarfArchives.org and maintained by Chris Gelino, J. Davy Kirkpatrick, and Adam Burgasser.  This research has made use of the Atomic Line List v2.04 maintained at \url{http://www.pa.uky.edu/$\sim$peter/atomic/}; the SIMBAD database and VizieR catalogue access tool, operated at CDS, Strasbourg, France; and the facilities of the Canadian Astronomy Data Centre operated by the National Research Council of Canada with the support of the Canadian Space Agency.  This paper uses data from the IRTF Spectral Library (\url{http://irtfweb.ifa.hawaii.edu/$\sim$spex/IRTF\_Spectral\_Library/index.html}) maintained by John Rayner, Michael Cushing, and William Vacca.  This publication makes use of data products from the Two Micron All Sky Survey, which is a joint project of the University of Massachusetts and the Infrared Processing and Analysis Center/California Institute of Technology, funded by the National Aeronautics and Space Administration and the National Science Foundation.  This research has made use of the NASA/IPAC Infrared Science Archive, which is operated by the Jet Propulsion Laboratory, California Institute of Technology, under contract with the National Aeronautics and Space Administration.  As some spectroscopic follow-up data was obtained from the summit of Mauna Kea, the authors wish to recognize and acknowledge the very significant and cultural role and reverence that this mountaintop has always had with the indigenous Hawaiian community.  We are most fortunate to have the opportunity to conduct observations there.

Facilities: \facility{IRTF~(SpeX), Magellan~(MagE), Blanco~(ISPI)}

\clearpage

\begin{deluxetable}{llllllll}
\tablewidth{6.5in}
\tablenum{1}
\tabletypesize{\scriptsize}
\tablecaption{Spectroscopic Observation Log for TWA 30}
\tablehead{
\colhead{Tel./Inst.} & \colhead{$\lambda$ ($\mu$m)} & \colhead{$\lambda$/$\Delta \lambda$} & 
\colhead{UT Date\tablenotemark{a}} & \colhead{N $\times$ t (s)\tablenotemark{b}} & 
\colhead{Z} & \colhead{Calibrator} & \colhead{Conditions}}
\startdata
Magellan/MagE & 0.30--1.05 & 4100 & 081126--1\tablenotemark{c} & 1~$\times$~300 & 1.4 & HR 3454 (WD) & Light Cirrus, 0$\farcs$5 seeing \\
Magellan/MagE & 0.30--1.05 & 4100 & 081126--2\tablenotemark{c} & 1~$\times$~300 & 1.4 & HR 3454 (WD) & Light Cirrus, 0$\farcs$5 seeing \\
Magellan/MagE & 0.30--1.05 & 4100 & 081126--3\tablenotemark{c} & 1~$\times$~300 & 1.4 & HR 3454 (WD) & Light Cirrus, 0$\farcs$5 seeing \\
Magellan/MagE & 0.30--1.05 & 4100 & 081127 & 1~$\times$~500 & 1.5 & GD 108 (WD) & Clear, 0$\farcs$6 seeing \\
Magellan/MagE & 0.30--1.05 & 4100 & 090106 & 1~$\times$~500 & 1.2 & GD 108 (WD) & Clear, 0$\farcs$8 seeing \\
Magellan/MagE & 0.30--1.05 & 4100 & 090110 & 1~$\times$~900 & 1.3 & GD 108 (WD) & Clear, 0$\farcs$5 seeing \\
Magellan/MagE & 0.30--1.05 & 4100 & 090111 & 1~$\times$~900 & 1.3 & GD 108 (WD) & Clear, 0$\farcs$5 seeing \\
Magellan/MagE & 0.30--1.05 & 4100 & 090305 & 1~$\times$~750 & 1.0 & GD 108 (WD) & Clear, 0$\farcs$5 seeing \\
Magellan/MagE & 0.30--1.05 & 4100 & 090531 & 1~$\times$~750 & 1.1 & EG 274 (WD) & Thin Clouds, 0$\farcs$9 seeing \\
\hline
Magellan/MIKE & 0.49--0.92 & 27000 & 081127 & 1~$\times$~500 & 1.7 & HR 3454 (WD) & Clear, 0$\farcs$6 seeing \\
\hline
IRTF/SpeX & 0.80--2.40 & 1200 & 081204 & 6~$\times$~120 & 1.7 & HD 92678 (A0 V) & Light Cirrus, 0$\farcs$6 seeing \\
IRTF/SpeX & 1.90--4.20 & 1500 & 081205 & 20~$\times$~30 & 1.8 & HD 92678 (A0 V) & 1--2 mags Cirrus, 0$\farcs$7 seeing \\
IRTF/SpeX & 0.80--2.40 & 1200 & 081215 & 12~$\times$~180 & 1.7 & HD 94741 (A0 V) & Clear, 0$\farcs$8 seeing \\
IRTF/SpeX & 0.80--2.40 & 1200 & 090202 & 10~$\times$~120 & 1.6 & HD 98949 (A0 V) & Light Cirrus, 0$\farcs$5 seeing \\
IRTF/SpeX & 0.80--2.40 & 1200 & 090514 & 6~$\times$~100 & 1.7 & HD 89911 (A0 V) & Clear, 0$\farcs$5 seeing \\
IRTF/SpeX & 0.80--2.40 & 1200 & 090515 & 8~$\times$~120 & 1.7 & HD 110653 (A0 V) & Clear, 0$\farcs$5 seeing \\
IRTF/SpeX & 0.80--2.40 & 1200 & 090520 & 8~$\times$~120 & 2.5 & HD 98949 (A0 V) & Clear, 0$\farcs$8 seeing \\
IRTF/SpeX & 0.80--2.40 & 1200 & 090616 & 6~$\times$~180 & 1.9 & HD 98949 (A0 V) & Light Cirrus, 0$\farcs$7 seeing \\
IRTF/SpeX & 0.70--2.50 & 150 & 090628 & 6~$\times$~10 & 2.1 & HD 98949 (A0 V) & Clear, 0$\farcs$5 seeing \\
IRTF/SpeX & 0.70--2.50 & 150 & 090629 & 6~$\times$~20 & 1.9 & HD 98949 (A0 V) & Clear, 0$\farcs$8 seeing \\
IRTF/SpeX & 0.80--2.40 & 1200 & 090629 & 4~$\times$~250 & 2.5 & HD 98949 (A0 V) & Clear, 0$\farcs$8 seeing \\
\enddata
\tablenotetext{a}{Epochs are denoted as YYMMDD in all tables.}
\tablenotetext{b}{Number of integrations times the integration time.}
\tablenotetext{c}{Three spectra were taken consecutively on 2008 Nov 26, which we denote as 1,2,3.}
\label{table:log}
\end{deluxetable}


\begin{deluxetable}{lll} 
\tablewidth{4.5in}
\tablenum{2}
\tabletypesize{\scriptsize}
\tablecaption{Properties of TWA 30}
\tablehead{ 
\colhead{Parameter} & \colhead{Value} & \colhead{Ref}}
\startdata
$\alpha$ (J2000)\tablenotemark{a} & 11 32 18.31 & 1 \\
$\delta$ (J2000)\tablenotemark{a} & $-$30 19 51.8 & 1 \\
\pmra & $-$89.6~$\pm$~1.3 mas yr$^{-1}$ & 2 \\
\pmdec & $-$25.8~$\pm$~1.3 mas yr$^{-1}$ & 2 \\
Distance\tablenotemark{b} & 42~$\pm$~2 pc & 3 \\
v$_{tan}$\tablenotemark{b} & 18.6~$\pm$~1.3 \kms & 3 \\
v$_{rad}$  & 12.3~$\pm$~1.5 \kms  & 3 \\
(\textit{U,V,W})\tablenotemark{b} & ($-$10.8,$-$19.2,$-$3.6)~$\pm$~(0.8,1.4,0.9) \kms & 3 \\
Optical SpT & M5~$\pm$~1 & 3 \\
Age\tablenotemark{c} & 7~$\pm$~3 Myr & 3 \\
Mass\tablenotemark{c} & 0.12~$\pm$~0.04 M$_{\sun}$ & 3 \\
log L$_{bol}$/L$_{\sun}$ & $-$1.64~$\pm$0.09 dex & 3 \\
$B\tablenotemark{d}$ & 15.6 & 4 \\
$R\tablenotemark{d}$ & 12.9 & 4 \\
$I\tablenotemark{d}$ & 11.30~$\pm$~0.03 & 5 \\
$J\tablenotemark{d}$ & 9.64~$\pm$~0.02 & 1 \\
$H\tablenotemark{d}$ & 9.03~$\pm$~0.02 & 1 \\
$K_s\tablenotemark{d}$ & 8.77~$\pm$~0.02 & 1 \\
Li EW\tablenotemark{e} & 0.61~$\pm$~0.13 \AA\ & 3 \\
H$\alpha$ EW\tablenotemark{f} & $-$6.8~$\pm$~1.2 \AA\ & 3 \\
X-ray & (2.5~$\pm$~1.0) $\times$ 10 $^{-2}$ counts s$^{-1}$ & 6 \\
HR1 & $-$0.2~$\pm$~0.4 & 6 \\
HR2 & 1.0~$\pm$~0.6 & 6 \\
log L$_X$/L$_{bol}$ & $-$3.34 dex & 3 \\
\enddata
\tablenotetext{a}{Coordinates are from the 2MASS Point Source Catalog at
epoch 1999 March 24 UT.}
\tablenotetext{b}{Estimated; see $\S$3.1.}
\tablenotetext{c}{See $\S$3.4.}
\tablenotetext{d}{Given in magnitudes.  TWA 30 has time variable photometry.  See references for further information.}
\tablenotetext{e}{Derived from the average equivalent widths measured from the seven MagE spectra reported in Table \ref{table:EWs_abs}.  The stated error is the 1$\sigma$ standard deviation in the measurement.}
\tablenotetext{f}{Derived from the average equivalent widths measured from the seven MagE spectra reported in Table \ref{table:EWs_balmer}.  The stated error is the 1$\sigma$ standard deviation in the measurement.}
\tablecomments{References: (1) 2MASS \citep{2006AJ....131.1163S}, (2) \cite{2010arXiv1002.0556F}, (3) this paper, 
(4) USNO-A2.0 \citep{1998yCat.1252....0M}, (5) DENIS \citep{1997Msngr..87...27E}, and (6) ROSAT All Sky Survey Faint Source Catalog \citep{2000IAUC.7432....3V}. 
}
\label{table:properties}
\end{deluxetable}

\clearpage

\begin{deluxetable}{lcc}
\tablewidth{3.5in}
\tablenum{3}
\tablecaption{Radial Velocity Measurements for TWA 30 from MagE Data}
\tablehead{ 
\colhead{UT Date} & \colhead{Abs. RV [km s$^{-1}$]\tablenotemark{a}} & 
\colhead{Rel. RV [km s$^{-1}$]\tablenotemark{b}}}
\startdata
081126$-$1 & 15.1 & N/A \\   
081126$-$2 & 12.5 & $-$1.1 \\    
081126$-$3 & 13.4 & $-$2.3 \\     
081127 & 11.3 & $-$4.0 \\    
090106 & 16.7 & 2.1 \\     
090110 & 13.5 & $-$1.8 \\     
090111 & 14.8 & 0.0 \\     
090305 & 5.9 & $-$8.2 \\     
090531 & 11.0 & $-$6.0 \\
\enddata
\tablenotetext{a}{These measurements were made by cross-correlation with radial velocity standards; see $\S$3.1.}
\tablenotetext{b}{These measurements were made by cross-referencing each spectrum to the 081126$-$1 UT spectrum and determining the relative offset.}
\label{table:rvs}
\end{deluxetable}

\begin{deluxetable}{llllllc}
\tablewidth{5.0in}
\tablenum{4}
\tabletypesize{\scriptsize}
\tablecaption{Equivalent Widths of Balmer Lines for TWA 30}
\tablehead{ 
\colhead{UT Date} & \colhead{$\lambda_{lab}$ (\AA)} & \colhead{$\lambda_{obs}$ (\AA)} & \colhead{Ion} & \colhead{Flux\tablenotemark{a}} & \colhead{EW$\pm$1$\sigma$ (\AA)} & \colhead{v\tablenotemark{b}}}
\startdata
081126$-$1 & 4102.892 & 4103.028 & H$\delta$ & 6.3 & $-$2.7~$\pm$~1.2 & 10 \\
081126$-$2 & 4102.892 & 4102.864 & H$\delta$ & 6.3 & $-$4.6~$\pm$~1.5 & $-$2 \\
081127 & 4102.892 & 4103.095 & H$\delta$ & 6.3 & $-$3.8~$\pm$~0.6 & 15 \\
090106 & \nodata & \nodata & H$\delta$ & \nodata & \nodata & \nodata \\
090110 & \nodata & \nodata & H$\delta$ & \nodata & \nodata & \nodata \\
090111 & \nodata & \nodata & H$\delta$ & \nodata & \nodata & \nodata \\
090305 & 4102.892 & 4103.013 & H$\delta$ & 4.9 & $-$3.1~$\pm$~0.4 & 9 \\
090531 & \nodata & \nodata & H$\delta$ & \nodata & \nodata & \nodata \\
\hline
081126$-$1 & 4341.684 & 4342.017 & H$\gamma$ & 10.0 & $-$6.1~$\pm$~1.0 & 23 \\
081126$-$2 & 4341.684 & 4341.914 & H$\gamma$ & 9.0 & $-$5.1~$\pm$~1.7 & 16 \\
081126$-$3 & 4341.684 & 4342.005 & H$\gamma$ & 9.6 & $-$5.8~$\pm$~2.4 & 22 \\
081127 & 4341.684 & 4341.933 & H$\gamma$ & 9.5 & $-$5.7~$\pm$~0.5 & 17 \\
090106 &\nodata & \nodata & H$\gamma$ & \nodata & \nodata & \nodata \\
090110 & 4341.684 & 4341.855 & H$\gamma$ & 9.4 & $-$5.5~$\pm$~1.8 & 12 \\
090111 & \nodata & \nodata & H$\gamma$ & \nodata & \nodata & \nodata \\
090305 & 4341.684 & 4341.796 &  H$\gamma$ & 6.9 & $-$4.0~$\pm$~0.2 & 8 \\
090531 & \nodata & \nodata & H$\gamma$ & \nodata & \nodata & \nodata \\
\hline
081126$-$1 & 4862.683 & 4862.951 & H$\beta$ & 10.3 & $-$6.7~$\pm$~0.3 & 17 \\
081126$-$2 & 4862.683 & 4862.915 & H$\beta$ & 10.2 & $-$6.7~$\pm$~0.2 & 14 \\
081126$-$3 & 4862.683 & 4862.909 & H$\beta$ & 12.6 & $-$8.9~$\pm$~1.3 & 14 \\
081127 & 4862.683 & 4862.838 & H$\beta$ & 11.2 & $-$7.7~$\pm$~0.2 & 10 \\
090106 & 4862.683 & 4863.037 & H$\beta$ & 11.5 & $-$7.5~$\pm$~2.6 & 22 \\
090110 & 4862.683 & 4862.863 & H$\beta$ & 10.2 & $-$6.6~$\pm$~0.7 & 11 \\
090111 & 4862.683 & 4862.871 & H$\beta$ & 9.6 & $-$6.0~$\pm$~1.2 & 12 \\
090305 & 4862.683 & 4862.823 & H$\beta$ & 9.4 & $-$5.9~$\pm$~0.1 & 9 \\
090531 & 4862.683 & 4862.816 & H$\beta$ & 8.4 & $-$4.9~$\pm$~0.2 & 8 \\
\hline
081126$-$1 & 6564.610 & 6564.940 & H$\alpha$ & 13.3 & $-$8.0~$\pm$~0.1 & 15 \\
081126$-$2 & 6564.610 & 6564.857 & H$\alpha$ & 12.9 & $-$7.6~$\pm$~0.1 & 11 \\
081126$-$3 & 6564.610 & 6564.799 & H$\alpha$ & 11.1 & $-$5.9~$\pm$~0.1 & 9 \\
081127 & 6564.610 & 6564.747 & H$\alpha$ & 12.9 & $-$8.1~$\pm$~0.1 & 6 \\
090106 & 6564.610 & 6564.811 & H$\alpha$ & 12.1 & $-$6.8~$\pm$~0.1 & 9 \\
090110 & 6564.610 & 6564.777 & H$\alpha$ & 12.1 & $-$6.8~$\pm$~0.1 & 8 \\
090111 & 6564.610 & 6564.854 & H$\alpha$ & 13.3 & $-$8.0~$\pm$~0.1 & 11 \\
090305 & 6564.610 & 6564.689 & H$\alpha$ & 9.5 & $-$4.6~$\pm$~0.1 & 4 \\
090531 & 6564.610 & 6564.647 & H$\alpha$ & 10.4 & $-$5.6~$\pm$~0.1 & 2 \\
\enddata
\tablenotetext{a}{The integrated line fluxes are given in units of 10$^{-16}$ erg cm$^{-2}$ s$^{-1}$ and should only be used to calculate relative line fluxes between features in the same spectrum as our data are not photometrically calibrated and hence do not account for slit losses or non-photometric conditions.}
\tablenotetext{b}{These velocities (\kms) are not corrected to the stellar rest frame (12.3~$\pm$~1.5 \kms).}
\label{table:EWs_balmer}
\end{deluxetable}

\clearpage 

\begin{deluxetable}{llllllc}
\tablewidth{5.0in}
\tablenum{5}
\tabletypesize{\scriptsize}
\tablecaption{Equivalent Widths of Selected Non-Balmer Emission Lines for TWA 30}
\tablehead{ 
\colhead{UT} & \colhead{$\lambda_{lab}$ (\AA)} & \colhead{$\lambda_{obs}$ (\AA)} & \colhead{Ion} & \colhead{Flux\tablenotemark{a}} & \colhead{EW$\pm$1$\sigma$ (\AA)\tablenotemark{b}} & \colhead{v\tablenotemark{c}}}
\startdata
081126$-$1 & 3727.090 & 3727.06 & [O II]\tablenotemark{d} & 11.5 & $-$8.5 & $-$2 \\
081126$-$2 & 3727.090 & 3727.05 & [O II]\tablenotemark{d} & 7.8 & $-$37.5 & $-$3 \\
081126$-$3 & 3727.090 & 3727.02 & [O II]\tablenotemark{d} & 7.1 & $-$24.9 & $-$6 \\
081127 & 3727.090 & 3726.94 & [O II]\tablenotemark{d} & 10.8 & $-$15.6 & $-$12 \\
090106 & 3727.090 & 3727.29 & [O II]\tablenotemark{d} & 4.2 & $-$17.1 & 16 \\
090110 & 3727.090 & 3727.08 & [O II]\tablenotemark{d} & 7.4 & $-$15.7 & $-$1 \\
090111 & 3727.090 & 3727.05 & [O II]\tablenotemark{d} & 4.1 & $-$58.9 & $-$3 \\
090305 & 3727.090 & 3727.09 & [O II]\tablenotemark{d} & 10.2 & $-$14.1 & 0 \\
090531 & 3727.090 & 3727.10 & [O II]\tablenotemark{d} & 3.7 & $-$30.2 & 1 \\
\hline
081126$-$1 & 3729.880 & 3729.91 & [O II]\tablenotemark{d} & 6.2 & $-$5.1 & 2 \\
081126$-$2 & 3729.880 & 3730.08 & [O II]\tablenotemark{d} & 4.9 & $-$14.5 & 16 \\
081126$-$3 & 3729.880 & 3729.88 & [O II]\tablenotemark{d} & 2.8 & $-$11.7 & 0 \\
081127 & 3729.880 & 3729.92 & [O II]\tablenotemark{d} & 6.7 & $-$9.8 & 3 \\
090106 & 3729.880 & 3729.83 & [O II]\tablenotemark{d} & 2.4 & $-$7.5 & $-$4 \\
090110 & 3729.880 & 3729.91 & [O II]\tablenotemark{d} & 4.6 & $-$6.8 & 2 \\
090111 & 3729.880 & 3729.65 & [O II]\tablenotemark{d} & 2.4 & $-$14.8 & $-$19 \\
090305 & 3729.880 & 3729.92 & [O II]\tablenotemark{d} & 6.0 & $-$9.7 & 3 \\
090531 & 3729.880 & 3729.88 & [O II]\tablenotemark{d} & 3.5 & $-$26.5 & 0 \\
\hline
081126$-$1 & 3934.777 & 3935.235 & Ca II k\tablenotemark{e} & 10.2 & $-$68.2 & 35 \\
$---$ & 3934.777 & 3933.775 & Ca II k\tablenotemark{e} & \nodata & \nodata & $-$76 \\
081126$-$2 & 3934.777 & 3935.271 & Ca II k\tablenotemark{e} & 34.0 & $-$29.7~$\pm$~6.9 & 38 \\
$---$ & 3934.777 & 3933.509 & Ca II k\tablenotemark{e} & \nodata & \nodata & $-$97 \\
081126$-$3 & 3934.777 & 3935.201 & Ca II k\tablenotemark{e} & 5.5 & $-$17.0 & 32 \\
$---$ & 3934.777 & \nodata & Ca II k\tablenotemark{e} & \nodata & \nodata & \nodata \\
081127 & 3934.777 & 3935.172 & Ca II k\tablenotemark{e} & 32.7 & $-$28.4~$\pm$~1.5 & 30 \\
$---$ & 3934.777 & 3933.389 & Ca II k\tablenotemark{e} & \nodata & \nodata & $-$106 \\
090106 & 3934.777 & 3935.14 & Ca II k\tablenotemark{e} & 4.0 & $-$50.7 & 34 \\
$---$ & 3934.777 & 3934.143 & Ca II k\tablenotemark{e} & \nodata & \nodata & $-$48 \\
090110 & 3934.777 & 3935.103 & Ca II k\tablenotemark{e} & 7.2 & $-$30.9 & 25 \\ 
$---$ & 3934.777 & 3933.953 & Ca II k\tablenotemark{e} & \nodata & \nodata & $-$63 \\
090111 & 3934.777 & 3935.02 & Ca II k\tablenotemark{e} & \nodata & $-$17.8 & 19 \\
$---$ & 3934.777 & 3933.799 & Ca II k\tablenotemark{e} & \nodata & \nodata & $-$75 \\
090305 & 3934.777 & 3935.204 & Ca II k\tablenotemark{e} & 34.8 & $-$30.2~$\pm$~1.5 & 33 \\
$---$ & 3934.777 & 3933.886 & Ca II k\tablenotemark{e} & \nodata & \nodata & $-$9 \\
090531 & 3934.777 & 3935.09 & Ca II k\tablenotemark{e} & 4.4 & $-$17.0 & 24 \\
081126$-$1 & 3969.591 & 3969.916 & Ca II h + H$\epsilon$\tablenotemark{d} & 17.3 & $-$12.2~$\pm$~2.3 & 25 \\
081126$-$2 & 3969.591 & 3969.71 & Ca II h + H$\epsilon$\tablenotemark{d} & \nodata & $-$6.1 & 26 \\
081126$-$3 & 3969.591 & 3970.100 & Ca II h + H$\epsilon$\tablenotemark{d} & 21.7 & $-$15.5~$\pm$~7.0 & 38 \\
081127 & 3969.591 & 3970.044 & Ca II h + H$\epsilon$\tablenotemark{d} & 28.4 & $-$23.3~$\pm$~2.8 & 34 \\
090106 & 3969.591 & 3969.946 & Ca II h + H$\epsilon$\tablenotemark{d} & 26.6 & $-$21.8~$\pm$~2.7 & 27 \\
090110 & 3969.591 & 3969.907 & Ca II h + H$\epsilon$\tablenotemark{d} & \nodata & $-$32.0 & 24 \\
090111 & 3969.591 & 3969.99 & Ca II h + H$\epsilon$\tablenotemark{d} & \nodata & $-$84.6 & 30 \\
090305 & 3969.591 & 3969.973 & Ca II h + H$\epsilon$\tablenotemark{d} & 29.0 & $-$24.0~$\pm$~2.0 & 29 \\
090531 & 3969.591 & 3969.85 & Ca II h + H$\epsilon$\tablenotemark{d} & \nodata & $-$33.0 & 29 \\
\hline
081126$-$1 & 4069.749 & 4069.918 & [S II] &19.3 & $-$15.4~$\pm$~2.0 & 13 \\
081126$-$2 & 4069.749 & 4069.829 & [S II] & 8.40 & $-$5.0~$\pm$~0.5 & 6 \\
081126$-$3 & 4069.749 & \nodata & [S II] & \nodata & \nodata & \nodata \\
081127 & 4069.749 & 4069.858 & [S II] & 13.5 & $-$10.1~$\pm$~0.6 & 8 \\
090106 & 4069.740 & 4069.68 & [S II] & 2.3 & $-$15.4 & $-$4 \\
090110 & 4069.749 & 4069.666 & [S II] & 11.8 & $-$7.9~$\pm$~1.6 & $-$6 \\
090111 & 4069.749 & \nodata & [S II] & \nodata & \nodata & \nodata \\
090305 & 4069.749 & 4069.802 & [S II] & 9.5 & $-$6.6~$\pm$~0.3 & 4 \\ 
090531 & 4069.749 & 4069.837 & [S II] & 8.9 & $-$6.3~$\pm$~2.6 & 7 \\
\hline
081126$-$1 & 4077.500 & \nodata & [S II] & \nodata & \nodata & \nodata \\
081126$-$2 & 4077.500 & \nodata  & [S II] & \nodata & \nodata & \nodata \\
081126$-$3 & 4077.500 & \nodata & [S II] & \nodata & \nodata & \nodata \\
081127 & 4077.500 & 4077.448 & [S II] & 5.2 & $-$2.5~$\pm$~0.4 & $-$4 \\
090106 & 4077.500 & 4077.57 & [S II] & $<$0.6 & $<-$2.7 & 5 \\
090110 & 4077.500 & \nodata & [S II] & \nodata & \nodata & \nodata \\
090111 & 4077.500 & \nodata & [S II] & \nodata & \nodata & \nodata \\
090305 & 4077.500 & 4077.558 & [S II] & 5.3 & $-$2.6~$\pm$~0.3 & 4\\
090531 & 4077.500 & \nodata & [S II] & \nodata & \nodata & \nodata \\
\hline
081126$-$1 & 4364.440 & 4364.966 & [O III] & 5.4 & $-$1.2~$\pm$~0.6 & 36 \\
081126$-$2 & 4364.440 & \nodata & [O III] & \nodata & \nodata & \nodata \\
081126$-$3 & 4364.440 & \nodata & [O III] & \nodata & \nodata & \nodata \\
081127 & 4364.440 & 4364.786 & [O III] & 6.2 & $-$2.1~$\pm$~0.4 & 24 \\
090106 & 4364.440 & \nodata & [O III] & \nodata & \nodata & \nodata \\
090110 & 4364.440 & \nodata & [O III] & \nodata & \nodata & \nodata \\
090111 & 4364.440 & \nodata & [O III] & \nodata & \nodata & \nodata \\
090305 & 4364.440 & 4364.770 & [O III] & 5.3 & $-$1.1~$\pm$~0.2 & 23 \\
090531 & 4364.440 & \nodata & [O III] & \nodata & \nodata & \nodata \\
\hline
081126$-$1 & 4572.377 & 4572.388 & Mg I] & 7.0 & $-$2.3~$\pm$~0.4 & 1 \\
081126$-$2 & 4572.377 & 4572.490 & Mg I] & 9.0 & $-$4.2~$\pm$~0.4 & 7 \\ 
081126$-$3 & 4572.377 & 4572.481 & Mg I] & 12.7 & $-$7.9~$\pm$~1.0 & 7 \\
081127 & 4572.377 & 4572.372 & Mg I] & 8.6 & $-$3.9~$\pm$~0.2 & 0 \\ 
090106 & 4572.377 & 4572.310 & Mg I] & 7.4 & $-$2.9~$\pm$~0.9 & $-$4 \\
090110 & 4572.377 & 4572.206 & Mg I] & 7.6 & $-$3.0~$\pm$~0.4 & $-$11 \\
090111 & 4572.377 & 4572.235 & Mg I] & 7.6 & $-$2.9~$\pm$~1.0 & $-$9 \\
090305 & 4572.377 & 4572.365 & Mg I] & 7.5 & $-$2.8~$\pm$~0.1 & $-$1 \\
090531 & 4572.377 & 4572.276 & Mg I] & 6.3 & $-$1.6~$\pm$~0.6 & $-$7 \\
\hline
081126$-$1 & 4659.350 & 4659.183 & [Fe III] & 5.1 & $-$1.7~$\pm$~0.2 & $-$11 \\  
081126$-$2 & 4659.350 & 4659.253 & [Fe III] & 5.1 & $-$1.3~$\pm$~0.2 & $-$6 \\
081126$-$3 & 4659.350 & \nodata & [Fe III] & \nodata & \nodata & \nodata \\
081127 & 4659.350 & 4659.500 & [Fe III] & 5.3 & $-$1.9~$\pm$~0.1 & 10 \\
090106 & 4659.350 & \nodata & [Fe III] & \nodata & \nodata & \nodata \\
090110 & 4659.350 & 4659.590 & [Fe III] & 5.4 & $-$1.7~$\pm$~0.3 & 15 \\
090111 & 4659.350 & \nodata & [Fe III] & \nodata & \nodata & \nodata \\
090305 & 4659.350 & 4659.684 & [Fe III] & 5.0 & $-$1.2~$\pm$~0.1 & 21 \\
090531 & 4659.350 & \nodata & [Fe III] & \nodata & \nodata & \nodata \\
\hline
081126$-$1 & 4756.020 & 4756.065 & [Fe III] & 2.3 & $-$0.2~$\pm$~0.1 & +3 \\
081126$-$2 & 4756.020 & \nodata & [Fe III] & \nodata & \nodata & \nodata \\
081126$-$3 & 4756.020 & \nodata & [Fe III] & \nodata & \nodata & \nodata \\
081127 & 4756.020 & 4756.789 & [Fe III] & 3.0 & $-$0.5~$\pm$~0.1 & 48 \\
090106 & 4756.020 & \nodata & [Fe III] & \nodata & \nodata & \nodata \\
090110 & 4756.020 & \nodata & [Fe III] & \nodata & \nodata & \nodata \\
090111 & 4756.020 & \nodata & [Fe III] & \nodata & \nodata & \nodata \\
090305 & 4756.020 & 4756.644 & [Fe III] & 2.5 & $-$0.4~$\pm$~0.1 & 39 \\
090531 & 4756.020 & \nodata & [Fe III] & \nodata & \nodata & \nodata \\
\hline
081126$-$1 & 4960.300 & 4960.380 & [O III] & 13.2 & $-$7.8~$\pm$~0.2 & 5 \\
081126$-$2 & 4960.300 & 4960.508 & [O III] & 15.1 & $-$9.4~$\pm$~0.2 & 13 \\
081126$-$3 & 4960.300 & 4960.452 & [O III] & 16.3 & $-$10.4~$\pm$~0.7 & 9 \\
081127 & 4960.300 & 4960.411 & [O III] & 12.7 & $-$7.2~$\pm$~0.1 & 7 \\
090106 & 4960.300 & 4960.389 & [O III] & 17.8 & $-$12.2~$\pm$~1.8 & 5 \\
090110 & 4960.300 & 4960.439 & [O III] & 17.9 & $-$12.4~$\pm$~1.0 & 8 \\
090111 & 4960.300 & 4960.303 & [O III] & 20.7 & $-$15.2~$\pm$~3.7 & 0 \\ 
090305 & 4960.300 & 4960.470 & [O III] & 10.0 & $-$4.1~$\pm$~0.1 & 10 \\
090531 & 4960.300 & 4960.570 & [O III] & 9.0 & $-$3.1~$\pm$~0.4 & 16 \\
\hline
081126$-$1 & 5008.240 & 5008.261 & [O III] & 33.6 & $-$25.1~$\pm$~0.2 & 1 \\
081126$-$2 & 5008.240 & 5008.277 & [O III] & 31.6 & $-$23.1~$\pm$~0.3 & 2 \\
081126$-$3 & 5008.240 & 5008.228 & [O III] & 36.4 & $-$27.9~$\pm$~0.6 & $-$1 \\
081127 & 5008.240 & 5008.219 & [O III] & 33.3 & $-$24.8~$\pm$~0.3 & $-$1 \\
090106 & 5008.240 & 5008.78 & [O III] & \nodata & $-$49.2 & 32 \\
090110 & 5008.240 & 5008.337 & [O III] & 34.2 & $-$25.6~$\pm$~0.7 & 6 \\
090111 & 5008.240 & 5008.313 & [O III] & 32.9 & $-$24.6~$\pm$~1.0 & 4 \\
090305 & 5008.240 & 5008.268 & [O III] & 18.9 & $-$10.5~$\pm$~0.1 & 2 \\
090531 & 5008.240 & 5008.299 & [O III] & 18.0 & $-$9.6~$\pm$~0.4 & 4 \\
\hline
081126$-$1 & 5877.227 & \nodata & He I &  \nodata & \nodata & \nodata \\
081126$-$2 & 5877.227 & 5876.947 & He I & 3.6 & $-$0.2~$\pm$~0.1 & $-$14 \\ 
081126$-$3 & 5877.227 & 5877.406 & He I & 4.0 & $-$0.6~$\pm$~0.1 & 9 \\
081127 & 5877.227 & 5877.451 & He I & 3.8 & $-$0.3~$\pm$~0.1 & 11 \\
090106 & 5877.227 & \nodata & He I &  \nodata & \nodata & \nodata \\
090110 & 5877.227 & \nodata & He I &  \nodata & \nodata & \nodata \\
090111 & 5877.227 & \nodata & He I &  \nodata & \nodata & \nodata \\
090305 & 5877.227 & 5877.130 & He I & 4.4 & $-$0.5~$\pm$~0.1 & $-$5 \\
090531 & 5877.227 & 5877.381 & He I & 3.6 & $-$0.2~$\pm$~0.1 & 8 \\
\hline
081126$-$1 & 5891.583 & 5892.387 & Na I D\tablenotemark{e} & 5.1 & $-$1.6~$\pm$~0.1 & 41 \\
$---$ & 5891.583 & 5890.280 & Na I D\tablenotemark{e} & \nodata & \nodata & $-$66 \\
081126$-$2 & 5891.583 & 5892.314 & Na I D\tablenotemark{e} & 5.3 & $-$1.9~$\pm$~0.1 & 37 \\ 
$---$ & 5891.583 & 5890.062 & Na I D\tablenotemark{e} & \nodata & \nodata & $-$77 \\
081126$-$3 & 5891.583 & 5892.286 & Na I D\tablenotemark{e} & 5.3 & $-$1.4~$\pm$~0.1 & 36 \\
$---$ & 5891.583 & 5890.193 & Na I D\tablenotemark{e} & \nodata & \nodata & $-$71 \\
081127 & 5891.583 & 5892.357 & Na I D\tablenotemark{e} & 5.5 & $-$2.0~$\pm$~0.1 & 39. \\
$---$ & 5891.583 & 5890.000 & Na I D\tablenotemark{e} & \nodata & \nodata & $-$81 \\
090106 & 5891.583 & 5892.210 & Na I D\tablenotemark{e} & 9.7 & $-$5.8~$\pm$~0.7 & 32 \\
$---$ & 5891.583 & 5890.436 & Na I D\tablenotemark{e} & \nodata & \nodata & $-$58 \\
090110 & 5891.583 & 5892.171 & Na I D\tablenotemark{e} & 6.4 & $-$3.3~$\pm$~0.2 & 30 \\
$---$ & 5891.583 & 5890.493 & Na I D\tablenotemark{e} & \nodata & \nodata & $-$56 \\
090111 & 5891.583 & 5892.245 & Na I D\tablenotemark{e} & 6.9 & $-$3.9~$\pm$~0.4 & 34 \\
$---$ & 5891.583 & 5890.496 & Na I D\tablenotemark{e} & \nodata & \nodata & $-$55 \\
090305 & 5891.583 & \nodata & Na I D\tablenotemark{f} & \nodata & \nodata & \nodata \\
090531 & 891.583 & \nodata & Na I D\tablenotemark{f} & \nodata & \nodata & \nodata \\
\hline
081126$-$1 & 5897.558 & 5898.417 & Na I D\tablenotemark{e} & 2.2 & $-$1.3 & 44\\
$---$ & 5897.558 & 5895.873 & Na I D\tablenotemark{e} & \nodata & \nodata & $-$86 \\
081126$-$2 & 5897.558 & 5898.333 & Na I D\tablenotemark{e} & 2.2 & $-$1.5 & 39 \\
$---$ & 5897.558 & 5896.308 & Na I D\tablenotemark{e} & \nodata & \nodata & $-$64 \\
081126$-$3 & 5897.558 & 5896.938 & Na I D\tablenotemark{e} & 5.0 & $-$1.1~$\pm$~0.1 & 36 \\ 
$---$ & 5897.558 & 5896.938 & Na I D\tablenotemark{e} & \nodata & \nodata & $-$32 \\
081127 & 5897.558 & 5898.398 & Na I D\tablenotemark{e} & 4.8 & $-$1.3~$\pm$~0.1 & 43 \\
$---$ & 5897.558 & 5896.095 & Na I D\tablenotemark{e} & \nodata & \nodata & $-$74 \\
090106 & 5897.558 & 5898.136 & Na I D\tablenotemark{e} & 7.4 & $-$4.1~$\pm$~0.7 & 29 \\
$---$ & 5897.558 & 5896.217 & Na I D\tablenotemark{e} & \nodata & \nodata & $-$98 \\
090110 & 5897.558 & 5898.106 & Na I D\tablenotemark{e} & 5.9 & $-$2.8~$\pm$~0.2 & 28 \\
$---$ & 5897.558 & 5896.452  & Na I D\tablenotemark{e} & \nodata & \nodata & $-$56 \\
090111 & 5897.558 & 5898.182 & Na I D\tablenotemark{e} & 6.1 & $-$3.0~$\pm$~0.4 & 32 \\
$---$ & 5897.558 & 5896.669 & Na I D\tablenotemark{e} & \nodata & \nodata & $-$45 \\
090305 & 5897.558 & \nodata & Na I D\tablenotemark{f} & \nodata & \nodata & \nodata \\
090531 & 5897.558 & \nodata & Na I D\tablenotemark{f} & \nodata & \nodata & \nodata \\
\hline
081126$-$1 & 6302.050 & 6302.223 & [O I] & 14.1 & $-$9.2~$\pm$~0.1 & 8 \\
081126$-$2 & 6302.050 & 6302.182 & [O I] & 13.5 & $-$8.9~$\pm$~0.1 & 6 \\ 
081126$-$3 & 6302.050 & 6302.206 & [O I] & 13.9 & $-$9.2~$\pm$~0.1 & 7 \\
081127 & 6302.050 & 6302.102 & [O I] & 14.5 & $-$9.9~$\pm$~0.1 & 2 \\
090106 & 6302.050 & 6302.068 & [O I] & 19.4 & $-$14.8~$\pm$~0.3 & 1 \\
090110 & 6302.050 & 6302.033 & [O I] & 18.3 & $-$13.7~$\pm$~0.1 & $-$1 \\
090111 & 6302.050 & 6302.114 & [O I] & 18.0 & $-$13.4~$\pm$~0.1 & 3 \\
090305 & 6302.050 & 6302.182 & [O I] & 7.8 & $-$3.2~$\pm$~0.1 & 6 \\
090531 & 6302.050 & 6302.313 & [O I] & 7.8 & $-$3.1~$\pm$~0.1 & 13 \\
\hline
081126$-$1 & 6365.540 & 6365.655 & [O I] & 6.3 & $-$2.1~$\pm$~0.1 & 5 \\
081126$-$2 & 6365.540 & 6365.721 & [O I] & 6.6 & $-$1.9~$\pm$~0.1 & 9 \\
081126$-$3 & 6365.540 & 6365.752 & [O I] & 6.8 & $-$2.2~$\pm$~0.1 & 10 \\
081127 & 6365.540 & 6365.530 & [O I] & 6.9 & $-$2.3~$\pm$~0.1 & $-$0 \\
090106 & 6365.540 & 6365.558 & [O I] & 7.6 & $-$3.4~$\pm$~0.1 & 1 \\
090110 & 6365.540 & 6365.507 & [O I] & 8.1 & $-$3.4~$\pm$~0.1 & $-$2 \\
090111 & 6365.540 & 6365.562 & [O I] & 7.4 & $-$3.2~$\pm$~0.1 & 1 \\
090305 & 6365.540 & 6365.717 & [O I] & 5.5 & $-$0.8~$\pm$~0.1 & 8 \\
090531 & 6365.540 & 6365.914 & [O I] & 5.5 & $-$0.9~$\pm$~0.1 & 18 \\
\hline
081126$-$1 & 6549.850 & 6550.278 & [N II] & 3.0 & $-$0.2~$\pm$~0.1 & 20 \\
081126$-$2 & 6549.850 & 6549.824 & [N II] & 3.1 & $-$0.2~$\pm$~0.1 & $-$1 \\
081126$-$3 & 6549.850 & \nodata & [N II] & \nodata & \nodata & \nodata \\
081127 & 6549.850 & 6550.045 & [N II] & 3.0 & $-$0.1~$\pm$~0.1 & 9 \\
090106 & 6549.850 & 6550.119 & [N II] & 3.2 & $-$0.4~$\pm$~0.1 & 12 \\
090110 & 6549.850 & 6549.875 & [N II] & 3.8 & $-$0.4~$\pm$~0.1 & 1 \\
090111 & 6549.850 & 6549.924 & [N II] & 3.7 & $-$0.3~$\pm$~0.1 & 3 \\
090305 & 6549.850 & 6549.632 & [N II] & 3.2 & $-$0.3~$\pm$~0.1 & $-$10 \\
090531 & 6549.850 & 6549.865 & [N II] & 3.1 & $-$0.2~$\pm$~0.1 & 1 \\
\hline
081126$-$1 & 6585.280 & 6585.279 & [N II] & 5.0 & $-$0.6~$\pm$~0.1 & 0 \\
081126$-$2 & 6585.280 & 6585.366 & [N II] & 5.4 & $-$0.6~$\pm$~0.1 & 4 \\
081126$-$3 & 6585.280 & 6585.170 & [N II] & 4.8 & $-$0.5~$\pm$~0.1 & $-$5 \\
081127 & 6585.280 & 6585.169 & [N II] & 4.9 & $-$0.5~$\pm$~0.1 & $-$5 \\
090106 & 6585.280 & 6585.407 & [N II] & 5.9 & $-$1.0~$\pm$~0.1 & 6 \\
090110 & 6585.280 & 6585.169 & [N II] & 5.1 & $-$0.8~$\pm$~0.1 & $-$5 \\
090111 & 6585.280 & 6585.377 & [N II] & 5.8 & $-$1.0~$\pm$~0.1 & 4 \\
090305 & 6585.280 & 6584.912 & [N II] & 5.1 & $-$0.3~$\pm$~0.1 & $-$17 \\
090531 & 6585.280 & 6585.067 & [N II] & 4.7 & $-$0.4~$\pm$~0.1 & $-$10 \\
\hline
081126$-$1 & 6718.290 & 6718.621 & [S II] & 3.9 & $-$0.9~$\pm$~0.1 & 15 \\
081126$-$2 & 6718.290 & 6718.504 & [S II] & 3.5 & $-$1.0~$\pm$~0.1 & 10 \\
081126$-$3 & 6718.290 & 6718.520 & [S II] & 3.7 & $-$1.2~$\pm$~0.1 & 10 \\
081127 & 6718.290 & 6718.518 & [S II] & 3.6 & $-$1.1~$\pm$~0.1 & 10 \\
090106 & 6718.290 & 6718.627 & [S II] & 5.5 & $-$2.6~$\pm$~0.1 & 15 \\
090110 & 6718.290 & 6718.521 & [S II] & 5.2 & $-$2.7~$\pm$~0.1 & 10 \\
090111 & 6718.290 & 6718.600 & [S II] & 5.5 & $-$2.5~$\pm$~0.1 & 14 \\
090305 & 6718.290 & 6718.376 & [S II] & 2.8 & $-$0.3~$\pm$~0.1 & 4 \\
090531 & 6718.290 & 6718.280 & [S II] & 3.0 & $-$0.5~$\pm$~0.1 & 0 \\
\hline
081126$-$1 & 6732.670 & \nodata & [S II] & \nodata & \nodata & \nodata \\
081126$-$2 & 6732.670 & 6733.000 & [S II] & 5.0 & $-$1.5~$\pm$~0.1 & 15 \\
081126$-$3 & 6732.670 & 6733.000 & [S II] & 5.0 & $-$1.6~$\pm$~0.1 & 15 \\
081127 & 6732.670 & 6732.888 & [S II] & 5.2 & $-$1.7~$\pm$~0.1 & 10 \\
090106 & 6732.670 & 6732.990 & [S II] & 7.5 & $-$4.0~$\pm$~0.1 & 14 \\
090110 & 6732.670 & 6732.913 & [S II] & 7.3 & $-$3.4~$\pm$~0.1 & 11 \\
090111 & 6732.670 & 6732.962 & [S II] & 6.9 & $-$3.5~$\pm$~0.1 & 13 \\
090305 & 6732.670 & 6733.000 & [S II] & 3.9 & $-$0.4~$\pm$~0.1 & 15 \\
090531 & 6732.670 & 6732.673 & [S II] & 4.1 & $-$0.6~$\pm$~0.1 & 0 \\
\hline
081126$-$1 & 7157.130 & 7157.077 & [Fe II] & 4.7 & $-$1.0~$\pm$~0.1 & $-$2 \\
081126$-$2 & 7157.130 & 7157.067 & [Fe II] & 4.7 & $-$1.0~$\pm$~0.1 & $-$3 \\
081126$-$3 & 7157.130 & 7157.235 & [Fe II] & 5.5 & $-$1.3~$\pm$~0.1 & 4 \\ 
081127 & 7157.130 & 7157.214 & [Fe II] & 5.5 & $-$1.3~$\pm$~0.1 & 4 \\
090106 & 7157.130 & 7157.215 & [Fe II] & 6.0 & $-$1.8~$\pm$~0.2 & 4 \\
090110 & 7157.130 & 7157.215 & [Fe II] & 6.1 & $-$2.0~$\pm$~0.1 & 4 \\
090111 & 7157.130 & 7157.357 & [Fe II] & 5.9 & $-$1.7~$\pm$~0.1 & 10 \\
090305 & 7157.130 & 7157.116 & [Fe II] & 4.3 & $-$0.6~$\pm$~0.1 & $-$1 \\
090531 & 7157.130 & 7157.088 & [Fe II] & 4.4 & $-$0.7~$\pm$~0.1 & $-$2 \\
\hline
081126$-$1 & 7173.980 & 7174.055 & [Fe II] & 3.9 & $-$0.3~$\pm$~0.1 & 3 \\ 
081126$-$2 & 7173.980 & 7173.893 & [Fe II] & 4.1 & $-$0.4~$\pm$~0.1 & $-$4 \\
081126$-$3 & 7173.980 & 7173.871 & [Fe II] & 4.2 & $-$0.5~$\pm$~0.1 & $-$5 \\
081127 & 7173.980 & 7174.041 & [Fe II] & 4.2 & $-$0.5~$\pm$~0.1 & 3 \\ 
090106 & 7173.980 & 7174.076 & [Fe II] & 4.4 & $-$0.7~$\pm$~0.1 & 4 \\
090110 & 7173.980 & 7173.923 & [Fe II] & 4.4 & $-$0.7~$\pm$~0.1 & $-$2 \\
090111 & 7173.980 & 7173.910 & [Fe II] & 4.3 & $-$0.6~$\pm$~0.1 & $-$3 \\
090305 & 7173.980 & 7173.874 & [Fe II] & 4.1 & $-$0.4~$\pm$~0.1 & $-$4 \\
090531 & 7173.980 & 7173.596 & [Fe II] & 3.5 & $-$0.3~$\pm$~0.1 & $-$16 \\
\hline
\enddata
\tablenotetext{a}{The integrated line fluxes are given in units of 10$^{-16}$ erg cm$^{-2}$ s$^{-1}$ and should only be used to calculate relative line fluxes between features in the same spectrum as our data are not photometrically calibrated and hence do not account for slit losses or non-photometric conditions.}
\tablenotetext{b}{Some lines were measured by hand and have typical uncertainties of 0.5--3 \AA.}
\tablenotetext{c}{These velocities (\kms) are not corrected to the stellar rest frame (12.3~$\pm$~1.5 \kms).}
\tablenotetext{d}{This line is blended or contaminated.}
\tablenotetext{e}{This line shows both blue- and red-shifted components.  The position of the weaker line component is reported but not the flux or EW.}
\tablenotetext{f}{This line has a p Cygni profile.}
\label{table:EWs_notbalmer}
\end{deluxetable}

\clearpage

\begin{deluxetable}{llllllc}
\tablewidth{5.0in}
\tablenum{6}
\tabletypesize{\scriptsize}
\tablecaption{Equivalent Widths of Selected Absorption Lines for TWA 30}
\tablehead{ 
\colhead{UT} & \colhead{$\lambda_{lab}$ (\AA)} & \colhead{$\lambda_{obs}$ (\AA)} & \colhead{Ion} & \colhead{Flux\tablenotemark{a}} & \colhead{EW$\pm$1$\sigma$ (\AA)} & \colhead{v\tablenotemark{b}}}
\startdata
081126$-$1 & 5330.014 & 5330.021 & Fe I & 1.8 & 0.6~$\pm$~0.1 & 0 \\
081126$-$2 & 5330.014 & \nodata & Fe I & \nodata & \nodata & \nodata \\
081126$-$3 & 5330.014 & 5329.688 & Fe I & 2.0 & 0.3~$\pm$~0.1 & $-$18 \\
081127 & 5330.014 & 5329.742 & Fe I & 2.2 & 0.5~$\pm$~0.1 & $-$15 \\
090106 & 5330.014 & \nodata & Fe I & \nodata & \nodata & \nodata \\
090110 & 5330.014 & 5329.991 & Fe I & 2.0 & 0.4~$\pm$~0.1 & $-$1 \\
090111 & 5330.014 & \nodata & Fe I & \nodata & \nodata & \nodata \\
090305 & 5330.014 & 5329.659 & Fe I & 1.8 & 0.6~$\pm$~0.1 & $-$20 \\ 
090531 & 5330.014 & 5329.668 & Fe I & 1.7 & 0.7~$\pm$~0.1 & $-$19 \\ 
\hline
081126$-$1 & 5372.983 & \nodata & Fe I & \nodata & \nodata & \nodata \\ 
081126$-$2 & 5372.983 & 5373.160 & Fe I & 1.7 & 0.3~$\pm$~0.1 & 10 \\
081126$-$3 & 5372.983 & 5373.200 & Fe I & 1.8 & 0.6~$\pm$~0.1 & 12 \\
081127 & 5372.983 & 5373.262 & Fe I & 1.9 & 0.5~$\pm$~0.1 & 16 \\
090106 & 5372.983 & \nodata & Fe I & \nodata & \nodata & \nodata \\ 
090110 & 5372.983 & \nodata & Fe I & \nodata & \nodata & \nodata \\ 
090111 & 5372.983 & \nodata & Fe I & \nodata & \nodata & \nodata \\ 
090305 & 5372.983 & 5373.091 & Fe I &1.6 & 0.4~$\pm$~0.1 & 6 \\ 
090531 & 5372.983 & 5373.176 & Fe I & 1.9 & 0.5~$\pm$~0.1 & 11 \\
\hline
081126$-$1 & 5398.629 & 5398.770 & Fe I & 1.6 & 0.4~$\pm$~0.1 & 8 \\
081126$-$2 & 5398.629 &  \nodata & Fe I & \nodata & \nodata & \nodata \\ 
081126$-$3 & 5398.629 &  \nodata & Fe I & \nodata & \nodata & \nodata \\ 
081127 & 5398.629 &  \nodata & Fe I & \nodata & \nodata & \nodata \\ 
090106 & 5398.629 &  \nodata & Fe I & \nodata & \nodata & \nodata \\ 
090110 & 5398.629 &  \nodata & Fe I & \nodata & \nodata & \nodata \\ 
090111 & 5398.629 & 5398.521 & Fe I & 1.9 & 0.5~$\pm$~0.1 & $-$6 \\
090305 & 5398.629 & 5398.597 & Fe I & 1.6 & 0.4~$\pm$~0.1 & $-$2 \\
090531 & 5398.629 &  \nodata & Fe I & \nodata & \nodata & \nodata \\ 
\hline
081126$-$1 & 5407.278 & \nodata & Fe I & \nodata & \nodata & \nodata \\  
081126$-$2 & 5407.278 & \nodata & Fe I & \nodata & \nodata & \nodata \\ 
081126$-$3 & 5407.278 & \nodata & Fe I & \nodata & \nodata & \nodata \\ 
081127 & 5407.278 & 5407.324 & Fe I & 2.1 & 0.3~$\pm$~0.1 & 3 \\ 
090106 & 5407.278 & \nodata & Fe I & \nodata & \nodata & \nodata \\ 
090110 & 5407.278 & 5407.136 & Fe I & 2.5 & 0.3~$\pm$~0.1 & $-$8 \\
090111 & 5407.278 & \nodata & Fe I & \nodata & \nodata & \nodata \\
090305 & 5407.278 & 5407.266 & Fe I & 2.0 & 0.4~$\pm$~0.1 & $-$1 \\
090531 & 5407.278 & 5407.266 & Fe I & 2.0 & 0.4~$\pm$~0.1 & $-$1 \\
\hline
081126$-$1 & 5411.276 & 5411.733 & Cr I & 2.5 & 0.3~$\pm$~0.1 & 25 \\
081126$-$2 & 5411.276 & 5411.443 & Cr I & 2.3 & 0.5~$\pm$~0.1 & 9 \\ 
081126$-$3 & 5411.276 & \nodata & Cr I & \nodata & \nodata & \nodata \\
081127 & 5411.276 & 5411.307 & Cr I & 2.4 & 0.3~$\pm$~0.1 & 2 \\
090106 & 5411.276 & \nodata & Cr I & \nodata & \nodata & \nodata \\
090110 & 5411.276 & 5411.713 & Cr I & 2.3 & 0.5~$\pm$~0.1 & 24 \\
090111 & 5411.276 & \nodata & Cr I & \nodata & \nodata & \nodata \\
090305 & 5411.276 & \nodata & Cr I & \nodata & \nodata & \nodata \\
090531 & 5411.276 & \nodata & Cr I & \nodata & \nodata & \nodata \\
\hline
081126$-$1 & 6123.912 & 6124.077 & Ca I & 2.6 & 0.5~$\pm$~0.1 & 8 \\
081126$-$2 & 6123.912 & 6123.961 & Ca I & 2.8 & 0.4~$\pm$~0.1 & 2 \\  
081126$-$3 & 6123.912 & 6124.128 & Ca I & 2.8 & 0.3~$\pm$~0.1 & 11 \\
081127 & 6123.912 & 6124.148 & Ca I & 2.8 & 0.3~$\pm$~0.1 & 12 \\
090106 & 6123.912 & \nodata & Ca I & \nodata & \nodata & \nodata \\
090110 & 6123.912 & 6123.958 & Ca I & 2.8 & 0.4~$\pm$~0.1 & 2 \\
090111 & 6123.912 & 6124.206 & Ca I & 2.7 & 0.4~$\pm$~0.1 & 14 \\ 
090305 & 6123.912 & 6123.981 & Ca I & 2.8 & 0.3~$\pm$~0.1 & 3 \\
090531 & 6123.912 & 6123.982 & Ca I & 2.7 & 0.4~$\pm$~0.1 & 3 \\
\hline
081126$-$1 & 6709.660 & 6710.010 & Li I & 3.8 & 0.6~$\pm$~0.1 & 16 \\ 
081126$-$2 & 6709.660 & 6709.928 & Li I & 3.7 & 0.7~$\pm$~0.1 & 12 \\
081126$-$3 & 6709.660 & 6709.953 & Li I & 3.8 & 0.6~$\pm$~0.1 & 13 \\
081127 & 6709.660 & 6709.794 & Li I & 3.9 & 0.6~$\pm$~0.1 & 6 \\
090106 & 6709.660 & 6709.946 & Li I & 3.0 & 0.5~$\pm$~0.1 & 13 \\ 
090110 & 6709.660 & 6709.913 & Li I & 3.0 & 0.4~$\pm$~0.1 & 11 \\
090111 & 6709.660 & 6709.995 & Li I & 2.9 & 0.5~$\pm$~0.1 & 15 \\
090305 & 6709.660 & 6709.753 & Li I & 3.1 & 0.8~$\pm$~0.1 & 4 \\
090531 & 6709.660 & 6709.769 & Li I & 3.2 & 0.8~$\pm$~0.1 & 5 \\
\hline
081126$-$1 & 7701.093 & 7701.091 & K I & 3.1 & 0.9~$\pm$~0.1 & 0 \\
081126$-$2 & 7701.093 & 7701.064 & K I & 3.1 & 0.9~$\pm$~0.1 & $-$1 \\
081126$-$3 & 7701.093 & 7701.015 & K I & 3.1 & 0.9~$\pm$~0.1 & $-$3 \\
081127 & 7701.093 & 7700.945 & K I & 3.3 & 0.7~$\pm$~0.1 & $-$6 \\ 
090106 & 7701.093 & 7701.001 & K I & 5.4 & $-$1.4~$\pm$~0.1\tablenotemark{c} & $-$4 \\
090110 & 7701.093 & 7700.869 & K I & 5.1 & $-$1.2~$\pm$~0.1\tablenotemark{c} & $-$9 \\
090111 & 7701.093 & 7700.895 & K I & 5.1 & $-$1.1~$\pm$~0.1\tablenotemark{c} & $-$8 \\
090305 & 7701.093 & 7701.001 & K I & 2.7 & 1.3~$\pm$~0.1 & $-$4 \\
090531 & 7701.093 & 7700.891 & K I & 2.5 & 1.4~$\pm$~0.1 & $-$8 \\
\hline
081126$-$1 & 8185.505 & 8185.689 & Na I & 3.2 & 1.1~$\pm$~0.1 & 7 \\
081126$-$2 & 8185.505 & 8185.622 & Na I & 3.3 & 0.9~$\pm$~0.1 & 4 \\
081126$-$3 & 8185.505 & 8185.611 & Na I & 3.2 & 1.0~$\pm$~0.1 & 4 \\
081127 & 8185.505 & 8185.532 & Na I & 3.8 & 1.0~$\pm$~0.1 & 1 \\
090106 & 8185.505 & 8185.748 & Na I & 3.2 & 1.0~$\pm$~0.1 & 9 \\
090110 & 8185.505 & 8185.533 & Na I & 3.8 & 1.0~$\pm$~0.1 & 1 \\
090111 & 8185.505 & 8185.732 & Na I & 3.3 & 1.0~$\pm$~0.1 & 8 \\
090305 & 8185.505 & 8185.439 & Na I & 3.9 & 0.9~$\pm$~0.1 & $-$2 \\
090531 & 8185.505 & 8185.633 & Na I & 3.3 & 0.9~$\pm$~0.1 & 5 \\
\enddata
\tablenotetext{a}{The integrated line fluxes are given in units of 10$^{-16}$ erg cm$^{-2}$ s$^{-1}$ and should only be used to calculate relative line fluxes between features in the same spectrum as our data are not photometrically calibrated and hence do not account for slit losses or non-photometric conditions.}
\tablenotetext{b}{These velocities (\kms) are not corrected to the stellar rest frame (12.3~$\pm$~1.5 \kms).}
\tablenotetext{c}{This line is seen in emission.}
\label{table:EWs_abs}
\end{deluxetable}

\clearpage

\begin{deluxetable}{llllllllll}
\tablewidth{5.0in}
\tablenum{7}
\tabletypesize{\footnotesize}
\tablecaption{Derived NIR Magnitudes\tablenotemark{a}, Colors\tablenotemark{a}, Spectral Types, and $A_v$ Values of TWA 30}
\tablehead{ 
\colhead{UT\tablenotemark{b}} & 
\colhead{$J$} & \colhead{$H$} & \colhead{$K_s$} & 
\colhead{$J-H$} & \colhead{$H-K_s$} & \colhead{$J-K_s$} & 
\colhead{NIR SpT\tablenotemark{c}} & \colhead{$A_v$} & 
\colhead{Notes}}
\startdata
990324 & 9.64  & 9.03 & 8.77   & 0.61  & 0.27 & 0.88 & \nodata & \nodata & 2MASS PSC \\
081204 & 10.2 & 9.3    & 8.8     & 0.91  & 0.43 & 1.34 & M5 & 2.5 & SpeX SXD \\
081215 & 10.5 & 9.6    & 9.2     & 0.85  & 0.41 & 1.25 & M5 & 2.2 & SpeX SXD \\
090202 &  9.8  & 9.0    & 8.7     & 0.75  & 0.35 &  1.09 & M4.5 & 1.5 & SpeX SXD \\
090514 & 11.3 & 10.2 & 9.6     & 1.09  & 0.61 &  1.71 & M5.5 & 4.2 & SpeX SXD \\
090515 & 11.7 & 10.5 & 9.7     & 1.23  & 0.75 &  1.98 & M6 & 5.6 & SpeX SXD \\
090520 & 12.7 & 11.0 & 10.0   & 1.68 &  0.95 &  2.63 & M6 & 9.0 & SpeX SXD \\
090616 & 10.6 & 9.6   & 9.2     & 0.91  & 0.47 &  1.39 & M5 & 2.6 & SpeX SXD \\
090628 & 10.6 & 9.7   & 9.2     & 0.96  & 0.49 &  1.45 & \nodata & \nodata & SpeX prism \\
090629 & 10.5 & 9.6   & 9.1     & 0.99  & 0.48 &  1.47 & \nodata & \nodata & SpeX prism \\
090629 & 10.5 & 9.5   & 9.1     & 0.93  & 0.48 &  1.41 & M5 & 2.7 & SpeX SXD \\
\enddata
\tablenotetext{a}{All magnitudes and colors are derived using 2MASS filters for spectrophotometry on the NIR spectra taken in SXD or prism mode (as indicated) with the exception of the epoch marked '2MASS PSC,' which is photometry.  In photometric conditions, the derived {\it absolute} spectrophotometry, flux calibrated with nearby A0 V stars, are accurate to $\sim$10\%, while the derived {\it relative} spectrophotometry (colors) are accurate to a few percent \citep{2009ApJS..185..289R}.   For a full description, see $\S$3.3.1.}
\tablenotetext{b}{UT dates are recorded as YYMMDD.}
\tablenotetext{c}{All NIR SpT and $A_v$ measurements were determined by dereddening the spectra to field M4-M8 templates and determining the best match; see $\S$3.3.1.}
\label{table:specphot}
\end{deluxetable}

\begin{deluxetable}{lcccccccc}
\tablewidth{6in}
\tablenum{8}
\tabletypesize{\scriptsize}
\tablecaption{Selected Values of MagE Optical Data for TWA 30}
\tablehead{ 
\colhead{UT Date\tablenotemark{a}} & \colhead{Opt SpT\tablenotemark{b}} & \colhead{A$_V$} & 
\colhead{H$\alpha$\tablenotemark{c}} & \colhead{[O I] $\lambda$6300\tablenotemark{c}} & 
\colhead{[N II] $\lambda$6583\tablenotemark{c}} & 
\colhead{[SII] $\lambda$6716\tablenotemark{c}} & \colhead{Li I $\lambda$6708\tablenotemark{c}} & 
\colhead{K I $\lambda$7699\tablenotemark{c}}}
\startdata
081126-1 & M4.75  & 1.6 & $-$8.0 & $-$9.2                         & $-$0.6 & $-$0.9 & 0.6 & 0.9  \\
081127    & M4.5     & 2.6 & $-$8.1 & $-$9.9                         & $-$0.5 & $-$1.1 & 0.6 & 0.7   \\
090106    & M4        & 0.6 & $-$6.8 & $-$14.8~$\pm$~0.3 & $-$1.0 & $-$2.6 & 0.5 & $-$1.4   \\
090110    & M4        & 0.3 & $-$6.8 & $-$13.7                      & $-$0.8 & $-$2.7 & 0.4 & $-$1.2  \\
090111    & M4        & 0.2 & $-$8.0 & $-$13.4                      & $-$1.0 & $-$2.5 & 0.5 & $-$1.1  \\
090305    & M5.25  & 1.7 & $-$4.6 & $-$3.2                         & $-$0.3 & $-$0.3 & 0.8 & 1.3  \\
090531    & M5.25  & 1.9 & $-$5.6 & $-$3.1                         & $-$0.4 & $-$0.5 & 0.8 & 1.4  \\
\enddata
\tablenotetext{a}{UT dates are recorded as YYMMDD.}
\tablenotetext{b}{These spectral types were determined from the dereddened data, described in $\S$3.2.1 and shown graphically in Figure \ref{fig:opt_spt_matches}.}
\tablenotetext{c}{The EW of the stated feature given in \AA.  The 1$\sigma$ uncertainty is 0.1 \AA\ unless otherwise noted.}
\label{table:AVs}
\end{deluxetable}

\clearpage

\clearpage

\bibliographystyle{apj}

\end{document}